\documentclass[fleqn, usenatbib]{mnras}
\usepackage{newtxtext,newtxmath}
\usepackage[T1]{fontenc}
\usepackage{ae,aecompl}
\usepackage{units}
\usepackage{graphicx}   
\usepackage{amsmath}    
\usepackage{amssymb}    
\usepackage{graphicx}   
\usepackage{amsmath}    
\usepackage{amssymb}    
\usepackage{layouts}
\usepackage{multicol}        
\usepackage{bm}         
\usepackage{pdflscape}  
\usepackage{verbatim}

\def\aap{A\&A}

\def\apj{ApJ}
\def\apjs{ApJS}

\def\mnras{MNRAS}

\def\pasp{PASP}

\def\araa{ARAA}

\title[Halo White Dwarfs]{The Age of the Galactic Stellar Halo from Gaia White Dwarfs}
\author[Kilic et al.]
{Mukremin Kilic$^1$,
P. Bergeron$^2$,
Kyra Dame$^{1}$,
N. C. Hambly$^3$,
N. Rowell$^3$, and
\newauthor
Courtney L. Crawford$^4$
\\
$^1$Homer L. Dodge Department of Physics and Astronomy, University of Oklahoma, 440 W. Brooks St., Norman, OK, 73019, USA\\
$^2$D\'epartement de Physique, Universit\'e de Montr\'eal, C.P. 6128, Succ. Centre-Ville, Montr\'eal, QC H3C 3J7, Canada\\
$^3$Institute for Astronomy, University of Edinburgh, Royal Observatory, Blackford Hill, Edinburgh EH9 3HJ, UK\\
$^4$Department of Physics and Astronomy, Louisiana State University, Baton Rouge, LA, 70803, USA\\
}

\date{\ \ Submitted \today \vspace{-0.5cm}}
\pubyear{2018}

\begin{document}
\label{firstpage}
\pagerange{\pageref{firstpage}--\pageref{lastpage}}
\maketitle

\begin{abstract}
We use 156 044 white dwarf candidates with $\geq5\sigma$ significant parallax measurements from the Gaia mission to measure the velocity
dispersion of the Galactic disc; $(\sigma_U,\sigma_V,\sigma_W) = (30.8, 23.9, 20.0)$ km s$^{-1}$. We identify 142 objects
that are inconsistent with disc membership at the $>5\sigma$ level. This is the largest sample of field halo white dwarfs
identified to date.  We perform a detailed model atmosphere analysis using optical and near-infrared photometry and parallaxes
to constrain the mass and cooling age of each white dwarf. The white dwarf cooling ages of our targets range from 7 Myr
for J1657+2056 to 10.3 Gyr for J1049$-$7400. The latter provides a firm lower limit of 10.3 Gyr for the age of the inner halo
based on the well-understood physics of white dwarfs. Including the pre-white dwarf evolutionary lifetimes, and limiting our
sample to the recently formed white dwarfs with cooling ages of $<500$ Myr, we estimate an age of $10.9 \pm 0.4$ Gyr
(internal errors only) for the Galactic inner halo. The coolest white dwarfs in our sample also give similar results. 
For example, J1049$-$7400 has a total age of 10.9-11.1 Gyr.  Our age measurements are consistent with other
measurements of the age of the inner halo, including the white dwarf based measurements of the globular clusters
M4, NGC 6397, and 47 Tuc.
\end{abstract}


\section{Introduction}

Age is a fundamental parameter for stars, yet the only fundamental age we can measure is that of the Sun through the decay
products of long-lived  isotopes in the solar system \citep{soderblom10}. There are a variety of model-dependent methods to
estimate the ages of the rest of the $\sim$200 billion stars in the Galaxy, but the main problem has been the lack of trigonometric
parallax measurements that enable a direct measurement of the luminosity of a star. Thanks to the Gaia Data Release 2
\citep{gaia18}, we now have an unprecedented opportunity to measure the luminosities and ages of a large number of stars in the Galactic disc and the stellar spheroid (hereafter we simply say halo).   

White dwarfs, with well-understood cooling physics, provide an excellent alternative to commonly used age indicators for the Galactic
disc and halo, including cluster main-sequence turn-off ages and nucleo-cosmochronometry \citep{sneden03,frebel07}. 
The cooling age of a white dwarf can be measured easily if the atmospheric composition, temperature, and radius (mass) are known.
White dwarfs have relatively simple atmospheres due to the high surface gravity, and the majority
of them display pure hydrogen atmospheres. 

There have been numerous efforts to measure the age of the Galactic disc using cool white dwarfs
\citep[e.g.,][]{winget87,liebert88,leggett98}, but nearby halo white dwarfs have been elusive.
Previous efforts to obtain a luminosity function for the halo have suffered from small numbers \citep{harris06,rowell11,kilic17},
except for the globular cluster white dwarfs in M4, NGC 6397, and 47 Tuc \citep{hansen04,hansen07,hansen13}. 

Most surveys for halo white dwarfs have relied on large area surveys done with photographic plates to find high proper motion objects, but they had limited success. For example,
\citet{monet00} searched 1378 square degrees and found one halo white dwarf candidate, POSS 15:00:03.51+36:00:30.5. WD 0343+247
\citep[also reported as WD 0346+246,][]{hambly99}, PM J13420$-$415 \citep{lepine05}, SDSS J110217.48+411315.4
\citep{hall08}, and LSR J0745+2627 \citep{catalan12} are other examples of high proper motion
(1.3, 2.55, 1.75, and $0\farcs9$ yr$^{-1}$, respectively) halo white dwarf candidates
found in such surveys. \citet{oppenheimer01} used the SuperCOSMOS sky survey \citep{hambly01} to identify 38 high 
proper motion halo white dwarf candidates. However, the majority of these white dwarfs are too warm and too young to belong to the halo
unless they are the descendants of sun-like stars with relatively long main-sequence lifetimes \citep{bergeron05}. Follow-up
ground-based parallax measurements for 15 of these stars confirmed halo membership at the $>3\sigma$ level for four
of them \citep{ducourant07}. 

\citet{kalirai12} identified four relatively warm ($T_{\rm eff}\geq14,000$ K) halo white dwarfs with $M\approx0.55 M_{\odot}$, and derived
an age of $11.4 \pm 0.7$ Gyr for the inner halo. Similarly, \citet{si17} used Bayesian hierarchical modeling to derive an age of $12.11^{+0.85}_{-0.86}$ Gyr based on 10 candidate halo white dwarfs, four of which had trigonometric parallax measurements available.
There are more than $10^5$ objects with significant parallaxes within the white dwarf region of the Gaia color-magnitude diagram. Here we use
this sample to measure the velocity dispersion of the Galactic disc and identify 142 objects as significant velocity outliers and as
members of the Galactic halo. Section 2 describes the Gaia sample selection and the velocity dispersion of the disc, whereas
Section 3 presents a detailed model atmosphere analysis of the halo white dwarfs based on optical and near-infrared photometry and
parallaxes. We present the results from this analysis in Section 4, discuss the total ages of these stars in Section 5, and
conclude in Section 6.

\section{The White Dwarf Sample}

\subsection{Gaia Sample Selection}

We queried the Gaia database for objects with $\varpi \geq 5 \sigma_{\varpi}$ significant parallaxes, and followed the recommendations
outlined in \cite{lindegren18} to remove non-Gaussian outliers in colour and absolute magnitude. We employed the
astrometric and photometric quality cuts outlined in Appendix~C of \cite{lindegren18}. We used a simple cut in
$(G_{\rm BP} - G_{\rm RP}, M_{\rm G})$ space keeping only those sources fainter than
the line joining (-1,5) and (5, 25) to identify the clearly subluminous stellar objects relative to the main sequence.
The query returned 156 044 sources.

\subsection{The Velocity Dispersion of the Disc}

\citet{fuhrmann11} use a sample of 284 stars with Hipparcos parallaxes to measure the mean space velocities and velocity
dispersions of $(U, V, W) = (2.5, -8.9, -1.0)$ km s$^{-1}$ and $(\sigma_U,\sigma_V,\sigma_W) = (30.8, 17.4, 14.0)$ km s$^{-1}$
for the thin disc. The majority of the white dwarfs identified in our query belong to the thin disc, but the fraction of thick disc white dwarfs 
is likely around 20\% \citep{reid05}. Instead of trying to isolate clean samples of thin and thick disc stars, we
treat the entire sample as one, since we are only interested in the significant velocity outliers. Since optical spectroscopy
and radial velocity observations are not available for the majority of these targets, we simply assume a zero radial velocity
and calculate the three dimensional velocities. Based on the 156 044 white dwarf candidates,
we measure $(U, V, W) = (4.8, -1.3, 3.0)$ km s$^{-1}$ and $(\sigma_U,\sigma_V,\sigma_W) = (30.8, 23.9, 20.0)$ km s$^{-1}$
for the Galactic disc. These are comparable to the mean velocities and velocity dispersions derived by \citet{fuhrmann11}. However,
the dispersions in $V$ and $W$ velocities are larger, likely due to the contribution from thick disc stars.

\subsection{Halo White Dwarfs}

\begin{figure}
\vspace{-0.2in}
\hspace{-0.1in}
\includegraphics[width=3.7in]{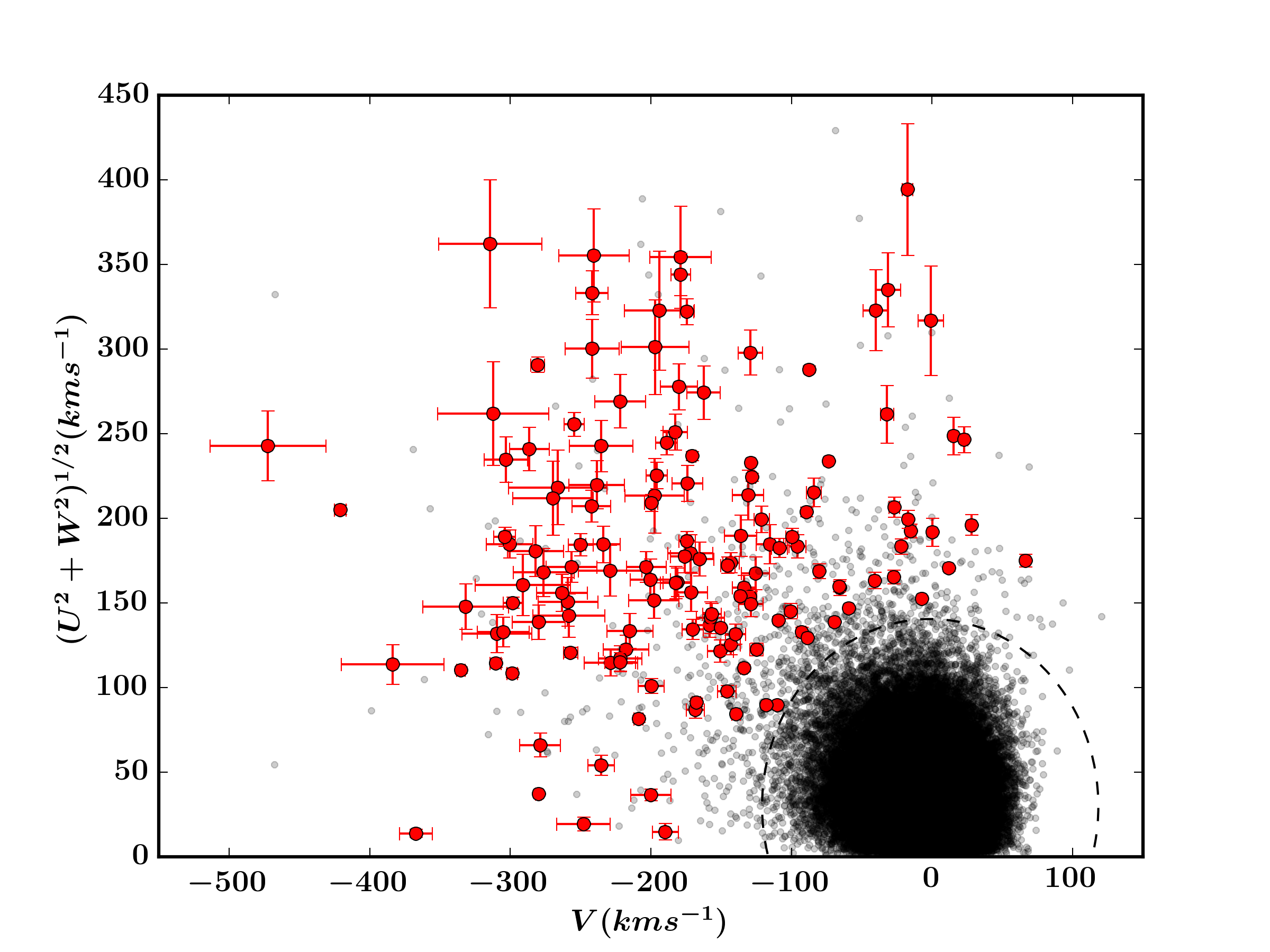}
\vspace{-0.2in}
\caption{Toomre diagram for 156 044 white dwarf candidates with $>5\sigma_{\varpi}$ significant parallax measurements in Gaia Data
Release 2 (open circles). The dashed line marks the $5\sigma_{UVW}$ velocity range of the entire sample.
Halo white dwarfs with velocities that are more than 5$\sigma$ away from this boundary are shown as red filled points.}
\label{fig:toom}
\end{figure}

Figure \ref{fig:toom} shows the Toomre diagram for our $\varpi \geq 5 \sigma_{\varpi}$ white dwarf sample. The dashed line marks the
$5\sigma_{UVW}$ velocity ellipsoid for the entire sample. We take a conservative approach, and select only those objects with
velocities that are more than
5$\sigma$ away from this boundary as members of the Galactic halo. There are 142 white dwarfs that are clearly not compatible with
a disc origin. As expected for a halo population, the majority of these velocity outliers lag behind the disc
with $V\sim-200$ km s$^{-1}$. 

\begin{table*}
\centering
\scriptsize
\caption{Gaia halo white dwarf sample. The positions are from Gaia Data Release 2, and are for the epoch 2015.5. Here we show the first five objects in the list. The entire sample is available as
a supplementary table in the online version of the article, which also includes all of the available optical and near-infrared
photometry.}
\begin{tabular}{cccccrrrcccc}
\hline
Gaia       & Name & Spectral & RA & DEC & $\varpi$ & $\mu_{\rm RA}$ & $\mu_{\rm DEC}$ & $G$ & $G_{BP}$ & $G_{RP}$ & $U, V, W$ \\
Source ID  &  & Type  & ($^{\circ}$) &  ($^{\circ}$) &   (mas)  & (mas yr$^{-1}$) & (mas yr$^{-1}$) & (mag) & (mag) & (mag) & (km s$^{-1}$) \\
\hline
2313582750735435776 & J000232.69$-$321149.1 & DA & 0.636201 & $-$32.196974 & 5.13 & 268.5 & $-$79.5 & 16.360 & 16.206 & 16.523 & $-$172.6, $-$165.3, $-$33.8 \\
2319735617804258176 & J000707.96$-$311337.4 & DB & 1.783158 & $-$31.227056 & 7.72 & 351.9 & $-$127.2 & 16.665 & 16.522 & 16.785 & $-$141.3, $-$156.7, $-$24.8 \\  
4922916739018189696 & J001311.26$-$550104.6 & \dots & 3.296902 & $-$55.017955 & 5.49 & 85.0 & $-$284.9 & 18.520 & 18.576 & 18.319 & 37.0, $-$221.9, 108.9 \\
2315640040069969152 & J003008.73$-$324402.9 & \dots & 7.536394 & $-$32.734135 & 4.26 & 269.9 & $-$60.1 & 19.292 & 19.334 & 19.141 & $-$213.0, $-$197.4, $-$11.1 \\
4703491116182416896 & J003428.89$-$684939.3 & \dots & 8.620368 & $-$68.827585 & 14.98 & $-$7.3 & $-$506.2 & 18.970 & 19.463 & 18.293 & 68.9, $-$92.8, 113.6 \\
\hline
\end{tabular}
\end{table*}

This sample includes two common-proper motion binary sytems (J075014.80+071121.3 plus J075015.56+071109.3, and
J115941.74$-$463034.3  plus J115956.86$-$462903.3), and several previously identified halo white dwarfs, including WD 0343+247 \citep{hambly99}, LSR J0745+2627 \citep{catalan12}, POSS 15:00:03.51+36:00:30.5 \citep{monet00}, WD 1448+077 and
WD 1524$-$749 \citep{kalirai12}, and several of the white dwarfs identified by \citet{oppenheimer01}. The proper motions
range from $0\farcs16$ yr$^{-1}$ to $3\farcs6$ yr$^{-1}$, with LHS 56 (WD 1756+827) being the highest proper motion halo
white dwarf identified here.

Out of the 142 white dwarfs, 59 have spectral classifications in the literature, including 35 DA, 1 DB, 20 DC, and 3 DQ
white dwarfs. Using Gaia positions and proper motions, we have cross-matched our halo white dwarf sample with the Sloan Digital Sky Survey Data Release 9 \citep{ahn12}, the Panoramic Survey Telescope and Rapid Response System
\citep[Pan-STARRS,][]{kaiser10}, the UKIRT Infrared Deep Sky Survey \citep[UKIDSS,][]{lawrence07}, the UKIRT
Hemisphere Survey \citep[UHS,][]{dye18}, and the VISTA Hemisphere Survey \citep{mcmahon13}. Most importantly, Pan-STARRS 3$\pi$ survey provides $grizy$ photometry
for 104 of our targets, and the SDSS provides $ugriz$ photometry for 52 of our halo white dwarfs. Table 1 presents the Gaia Source
identifications, spectral types, positions, parallaxes, proper motions, photometry, and $UVW$ space velocities for each
target. The full table is available as supplementary information in the online version of the article and includes all of the
available optical and near-infrared photometry for each white dwarf.

\begin{figure}
\vspace{-0.5in}
\hspace{-0.1in}
\includegraphics[width=3.7in]{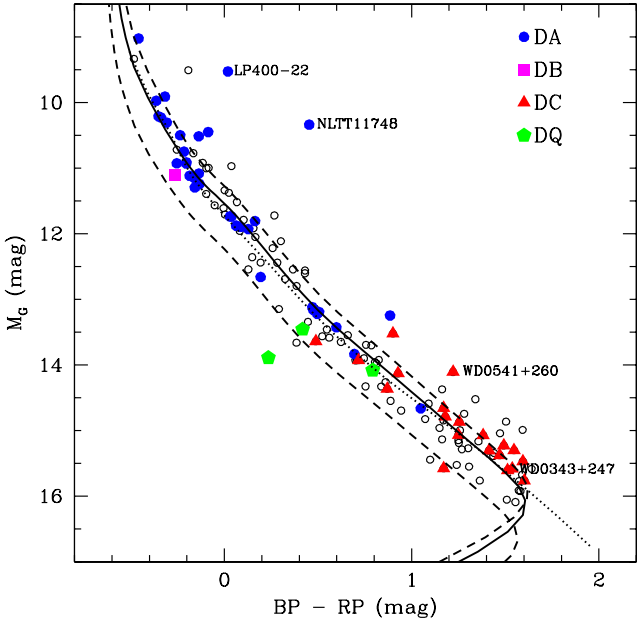}
\vspace{-0.8in}
\caption{Gaia color-magnitude diagram for the halo white dwarf sample. The solid and dotted lines show the cooling sequences
for $0.5 M_{\odot}$ pure H and pure He atmosphere white dwarfs, respectively, whereas the dashed lines show the same
sequences for 0.4 and 0.8 $M_{\odot}$ pure H atmosphere white dwarfs. Colored symbols mark objects with optical spectra. 
Two ELM white dwarfs, LP400$-$22 and NLTT 11748, and two DC white dwarfs with mixed H/He atmospheres, WD 0343+247 and WD 0541+260 are labelled.}
\label{fig:gaia}
\end{figure}

\begin{figure*}
\vspace{-0.5in}
\includegraphics[width=3.6in]{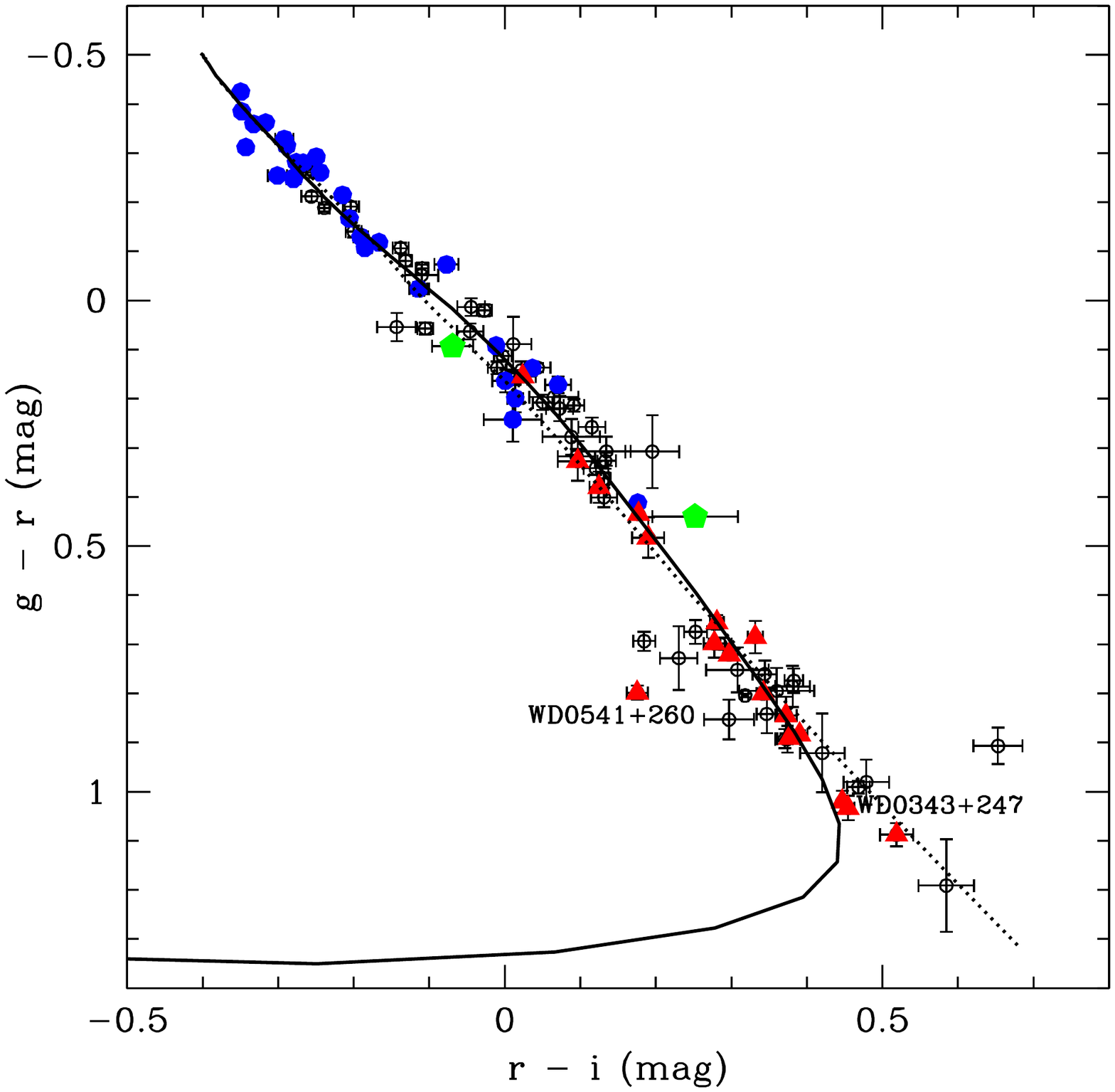}
\hspace{-0.3in}\includegraphics[width=3.6in]{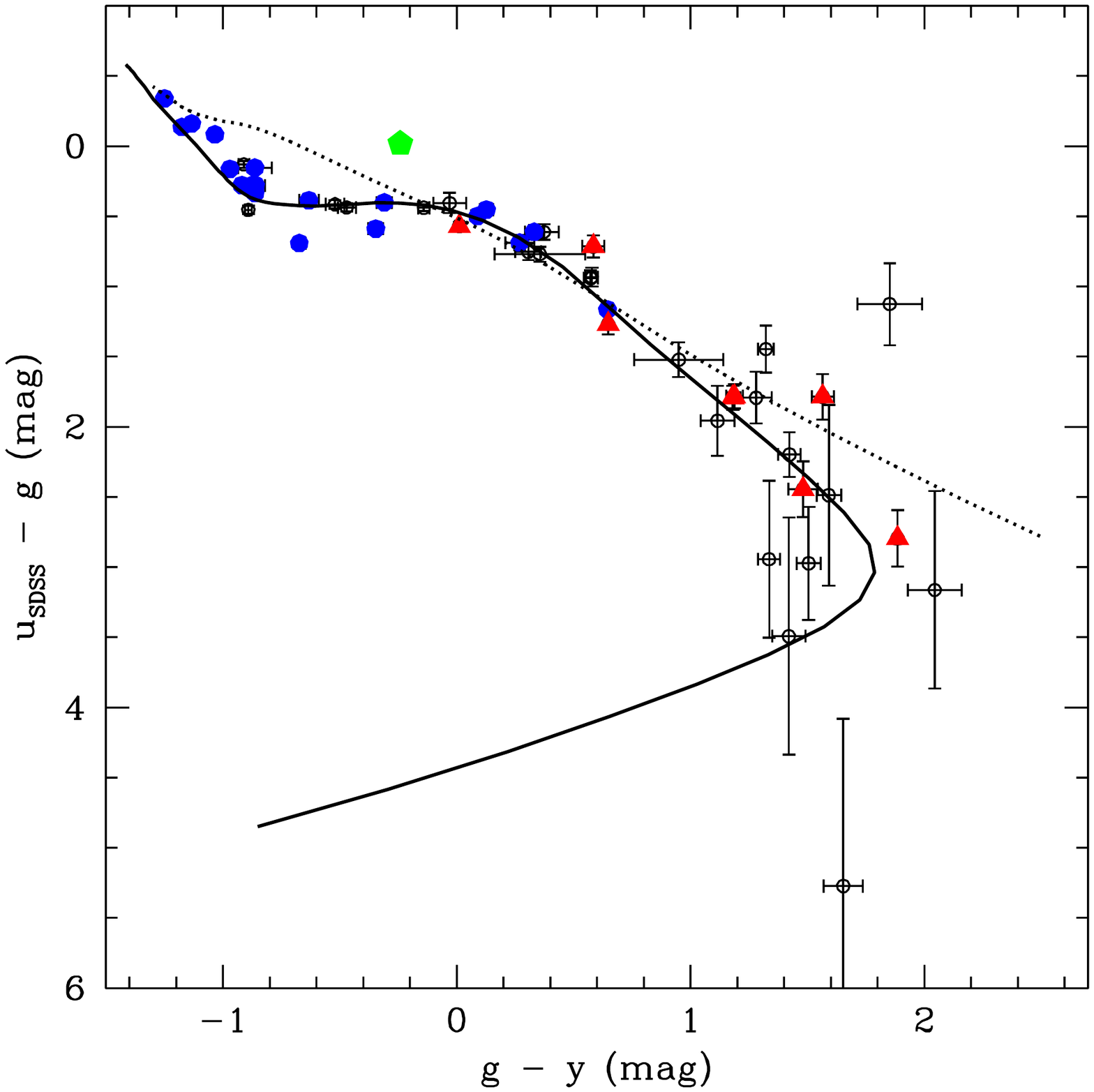}
\vspace{-0.8in}
\caption{Colour-colour diagrams for our halo white dwarf sample based on the Pan-STARRS $grizy$ and SDSS $u-$band
photometry. The symbols and cooling sequences are the same as in Fig. \ref{fig:gaia}.}
\label{fig:pan}
\end{figure*}

Figure \ref{fig:gaia} presents the Gaia color-magnitude diagram for our halo white dwarf sample, along with the cooling
tracks for $0.5 M_{\odot}$ pure H and pure He atmosphere white dwarfs (solid and dotted lines, respectively). To
illustrate the mass range of our sample of white dwarfs, we also plot the cooling sequences for 0.4 and $0.8 M_{\odot}$
pure H atmosphere white dwarfs as dashed lines. The pure H atmosphere
tracks show the blue hook due to the collision induced absorption from molecular hydrogen for $M_G\geq16$ mag
and $T_{\rm eff}\leq3500$ K. Objects with spectral classifications are shown with colored symbols. Interestingly,
this sample includes objects with a range of colors (temperatures) and magnitudes, and therefore a range of white dwarf
masses and cooling ages. Two of these white dwarfs, LP 400-22 and NLTT 11748 (labelled in the figure), belong to the
Extremely Low Mass (ELM, $M\leq0.3M_{\odot}$) white dwarf population and they are both found in short period binary systems \citep{kilic09,kilic10,kawka10,steinfadt10}. Two other white dwarfs with mixed H/He atmospheres (see below), WD 0343+247 and
WD 0541+260, are also labelled.

Most of the white dwarfs in this figure lie near the $0.5 M_{\odot}$ tracks, indicating that they are relatively low-mass.
A $\sim10$ Gyr old population should be currently forming $\sim0.5M_{\odot}$ white dwarfs. Hence, the overabundance
of relatively low-mass white dwarfs at the bright end of the cooling sequence is not surprising. Even though the faintest (and
hence oldest) white dwarfs are expected to be on average more massive than those near the top of the cooling sequence
(as they should have evolved from more massive stars), there are theoretical considerations that make their detections unlikely.
 
 Analyzing the local sample of white dwarfs with parallax measurements, \citet{fontaine01} and \citet{blr01} demonstrate
 that at low temperatures, massive white dwarfs would be indeed very old, but they are rare because of the relative
 under-abundance of their progenitor massive stars. More importantly,
 the oldest massive white dwarfs would have entered the rapid Debye cooling phase \citep{vanhorn68} and disappeared from observational
 samples. This is also confirmed through the observations of the faintest white dwarfs in globular clusters.
 For example, Figure 19 of \citet{hansen07} shows the white dwarf luminosity function and the mass distribution
 for the globular cluster NGC 6397. \citet{hansen07} found that the bulk of the luminosity function contains white dwarfs
 of the same mass ($0.52 M_{\odot}$) and that the masses start to increase to $0.62 M_{\odot}$ at the
 truncation of the luminosity function. Hence, they conclude that the truncation in the luminosity function occurs due to the faster cooling timescales
 of more massive white dwarfs at late times. Figure 8 of \citet{fontaine01} shows the theoretical isochrones with and
 without the main-sequence lifetimes taken into account. Because of the particular S-shape of the isochrones, the majority
 of the oldest white dwarfs are actually found close to the $0.6 M_{\odot}$ tracks. Hence, the relatively large numbers of
 $\sim0.5 M_{\odot}$ cool white dwarfs in our sample is consistent with the theoretical expectations as well as the observed
 globular cluster white dwarf cooling sequences.

Figure \ref{fig:pan} shows colour-colour diagrams for our sample of halo white dwarfs that are also detected in Pan-STARRS (left panel)
and the SDSS (right panel). 
The symbols and cooling tracks are the same as in Fig \ref{fig:gaia}. The colours for the observed sample follow the predictions
from the pure H or pure He atmosphere models relatively well, and the coolest white dwarfs seem to have temperatures near 3500 K.
In addition, several WDs with $u-g \approx 0.5$ mag and unknown spectral types follow the pure H atmosphere tracks, indicating that
they suffer from the Balmer jump in the $u-$band due to a H-rich atmosphere.

\section{Model Atmosphere Analysis}

We use the photometric technique described at length in \citet{bergeron97}. Briefly, we convert the available
magnitudes $m$ into average fluxes, $f^{m}_{\lambda}$, using a procedure similar to that outlined in \citet{holberg06}.
The SDSS and Pan-STARRS are on the AB magnitude system, while the other datasets, including Gaia, are on the
Vega system.  After estimating the average fluxes, we compare them with the model Eddington fluxes,
$H^{m}_{\lambda}$, properly averaged over the appropriate filter bandpass. These two average fluxes are related by the equation
\begin{equation}
f^{m}_{\lambda} = 4\pi(R/D)^2H^{m}_{\lambda}
\end{equation}

\noindent where $R/D$ defines the ratio of the radius of the star to its distance from Earth. We define a $\chi^2$ value in
terms of the difference between observed and model fluxes over all bandpasses, properly weighted by the photometric
uncertainties, and minimize it using the nonlinear least-squares method of Levenberg-Marquardt \citep{press86}.
We treat both $T_{\rm eff}$ and the solid angle $\pi(R/D)^2$ as free parameters, perform fits for both pure H and pure He
atmospheres, and obtain the uncertainties for each fitted parameter directly from the covariance matrix of the fitting
algorithm. Using Gaia parallaxes, we obtain the radius $R$ directly from the solid angle and the distance $D$, and use
the evolutionary models for C/O white dwarfs \citep{blr01} to estimate the mass and cooling age of each object.

Only 55 of our halo white dwarfs are within 100 pc of the Sun. Since they are within the Local Bubble \citep{lallement03}, we assume that
interstellar extinction is negligible for these stars. However, the remaining 87 targets suffer from interstellar extinction.  
We follow \citet{harris06} and correct for full extinction for stars beyond 250 pc, and partial extinction, which changes linearly with
distance, for stars between 100 pc and 250 pc. We use the $E(B-V)$ values from \citet{schlafly11}, and the extinction coefficients of
$R=3.384$, 2.483, 1.838, 1.414, 1.126, 0.650, 0.327, and 0.161 for the Pan-STARRS $grizy$ and the near-infrared
$JHK$ filters, respectively \citep{green18}. We use the extinction coefficients from \citet{gentile18} for Gaia photometry.

We present the SEDs and our model fits to both observed and dereddened photometry for all 142 targets as supplementary
data in the online version of the article. The median Galactic latitude $\lvert l \rvert$ and $E(B-V)$ for the 87 stars beyond 100 pc are
$39^{\circ}$ and 0.03 mag, respectively. This sample includes three stars within $10^{\circ}$ of the Galactic plane:
J053712.48+321501.0, J091304.64-375131.0, and J214959.88+540841.9. Two of these are spectroscopically confirmed DA
white dwarfs.  All three have $T_{\rm eff}\sim$10 000 K, $d =100$-200 pc, and our best-fit solutions change by up to 190 K in
temperature and 0.02 dex in $\log{g}$ after the extinction correction. These have negligible effects on our cooling age estimates. 

Our spectral energy distribution (SED) fits using SDSS $ugriz$, Pan-STARRS $grizy$, Gaia $G, G_{\rm BP}, G_{RP}$, 
and JHK photometry show that the quality of the Pan-STARRS $grizy$ photometry is superior to the SDSS photometry, and
the addition of the Pan-STARRS $y$ filter photometry helps constrain the SED fits much better, especially for cool white dwarfs. We also
find that given the broad passbands, the Gaia photometry is not as useful in constraining our model fits. Therefore, we use
the Pan-STARRS $grizy$ and near-infrared $JHK$ photometry for the 104 targets that have Pan-STARRS data available,
plus the SDSS $u$-band data, if available. The $u$-band filter helps with distinguishing H atmosphere white dwarfs since
it covers the Balmer jump. For the rest of the targets, we rely on the SDSS and/or Gaia photometry plus $JHK$ filters to constrain the fits.

\begin{figure*}
\vspace{-1in}
\includegraphics[angle=-90,width=7in]{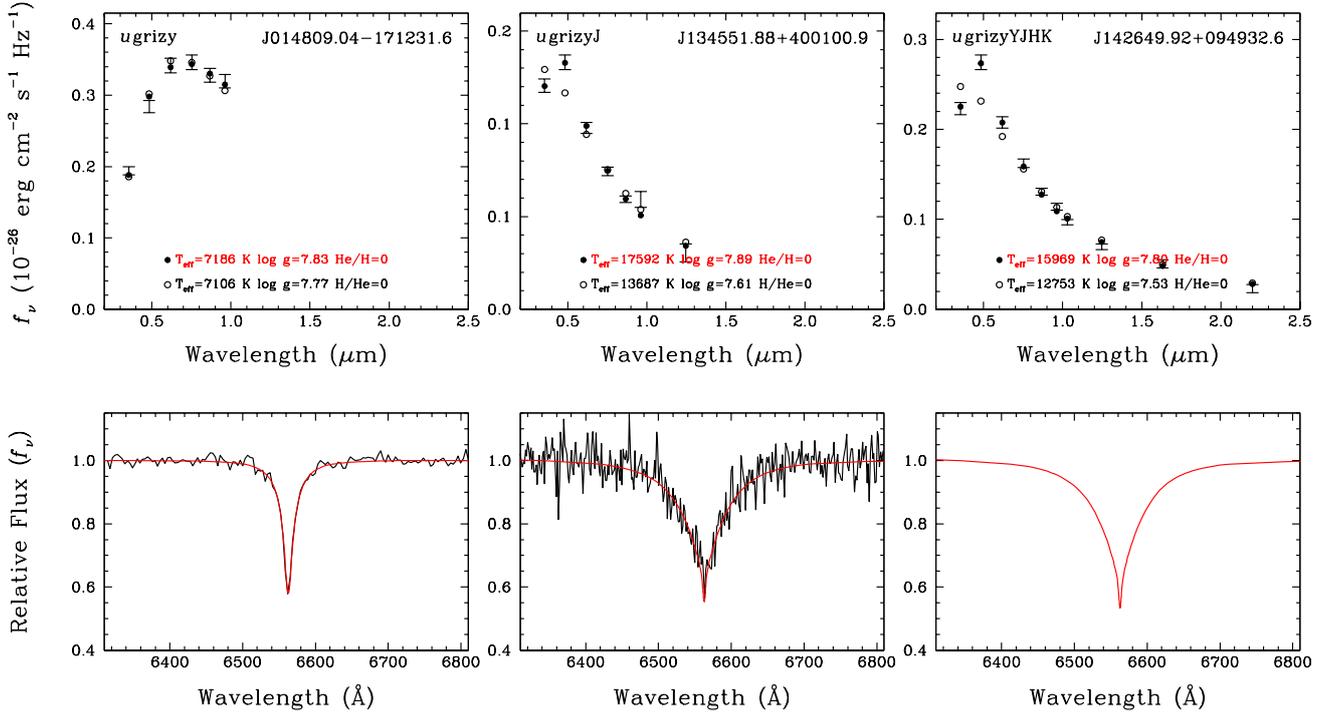}
\vspace{-0.5in}
\caption{{\it Top panels:} SED fits for three white dwarfs that are best explained by pure H atmosphere white dwarf models. 
Error bars show the data, and the filled and open circles show the predictions from
the best-fitting pure H and pure He atmosphere white dwarf models, respectively. 
{\it Bottom Panels:}  Optical spectra (black lines, if available) around the H$\alpha$ region for the same stars,
and the predicted H$\alpha$ line profile from the best-fitting pure H atmosphere white dwarf model (red lines).}
\label{fig:da}
\end{figure*}

\begin{figure*}
\vspace{-1in}
\includegraphics[angle=-90,width=7in]{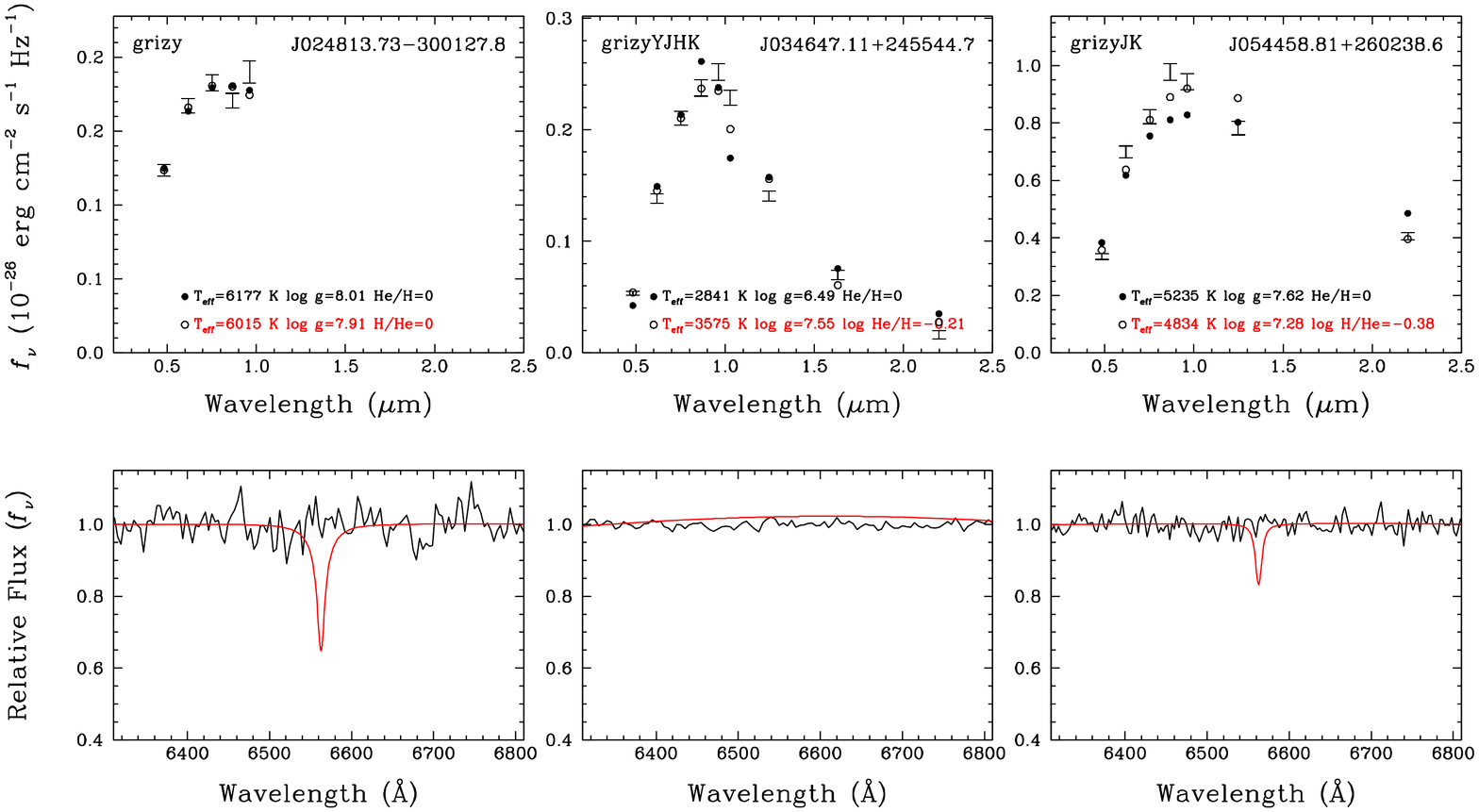}
\vspace{-0.5in}
\caption{SED fits for three DC white dwarfs with He-rich atmospheres. The symbols are the same as in Fig \ref{fig:da}.
The bottom panels show the optical spectra (black lines) around the H$\alpha$ region for the same stars, and
the predicted H$\alpha$ line profiles from the best-fitting pure H atmosphere white dwarf models. The lack of H$\alpha$
absorption clearly indicates He-rich atmospheres for J0248$-$3001 and J0544+2602, and
the observed photometry favors mixed atmospheres with roughly equal amounts of H and He in 
J0346+2455 (WD 0343+247) and J0544+2602.}
\label{fig:dc}
\end{figure*}

\citet{gentile18} compared the best-fit atmospheric parameters for nearly 5,000 DA white dwarfs based on the Pan-STARRS, SDSS, 
and Gaia photometry (see their Figures 7 and 8), and found that they are in good agreement across the temperature range
5,000 - 50,000 K with no clear systematic trends. \citet{hollands18} studied the 20 pc local white dwarf sample from Gaia, and found standard
deviations of 3.1\% in temperature and 0.1 dex in $\log{g}$ in the published parameters.  

Comparing the model fits for our halo
white dwarf sample based on Pan-STARRS and SDSS photometry, we find average differences of $0.8 \pm 2.3$\% in
temperature and $0.005 \pm  0.053 M_{\odot}$ in mass, respectively. These are well within the typical uncertainties of our measurements. J115131.13+015952.6, is the only significant outlier in this sample, however given its relatively cool temperature (4540-4580 K), its spectral energy distribution peaks around $1\mu$m, and Pan-STARRS $z-$ and $y-$band photometry provide superior constraints on its temperature compared to the SDSS. Similarly, comparing the model fits based on Pan-STARRS and
Gaia photometry, we find average differences of $1.6 \pm 4.7$\% in temperature and $0.006 \pm 0.062 M_{\odot}$ in mass,
respectively. Again, these are well within the typical uncertainties of our measurements, and there is no evidence of significant systematic effects. There are 11 objects with significant mass differences of $>0.1 M_{\odot}$. However, the majority of these objects
are cooler than 5,000 K, and therefore Pan-STARRS $z-$ and $y-$band photometry provide superior constraints on temperature
(and therefore mass), since temperatures and radii (and masses) of the best-fit models are correlated. 

\section{Results}

\subsection{DA White Dwarfs}

There are 35 spectroscopically confirmed DA white dwarfs in our sample, and we use the pure H atmosphere models to fit their SEDs.
Figure \ref{fig:da} shows our SED fits  (top panels) to three white dwarfs that are best-explained as pure H atmosphere white dwarfs.
The error bars show the data, and the filled and open circles show the predictions from the best-fitting
pure H and pure He atmosphere models. The bottom panels show the optical spectra \citep[if available in the Montreal
White Dwarf Database,][]{dufour17} along with the predicted H$\alpha$ lines for the best-fitting pure H
atmosphere models to the photometry. Note that this is not a fit to the spectrum, but just a comparison to make sure that
our photometric solution is consistent with the observed H$\alpha$ line profiles for the white dwarfs that have optical spectra available. 

Figure \ref{fig:da} shows that our photometric solutions for both J0148$-$1712 and J1345+4001 provide an excellent match to
the H$\alpha$ line profiles for these stars, indicating that these are clearly pure H atmosphere white dwarfs. J1345+4001 is warm enough
to show a strong Balmer jump in the $u$-band, and its SED demonstrates the usefulness of the $u$-band photometry
to identify H atmosphere white dwarfs that lack follow-up spectroscopy. The right panels show our fits to one such white dwarf, J1426+0949,
that lacks optical spectroscopy. The SED for J1426+0949 shows a significant Balmer jump in the $u$-band, clearly
demonstrating that this is a H atmosphere white dwarf. We inspect the fits for all 52 targets with SDSS $u-$band data available, and
identify additional H atmosphere white dwarfs that lack follow-up optical spectroscopy, but show the Balmer jump. 

\subsection{DB and DQ White Dwarfs}

There are 1 DB and 3 DQ white dwarfs in our halo white dwarf sample. We use pure He atmosphere models to fit the SED for the
DB white dwarf J0007$-$3113, and He atmosphere models with trace amounts of C to fit both the SEDs and the optical spectra
for the DQ white dwarfs \citep[see also][]{dufour05,giammichele12}. One of the DQ white dwarfs, J1045$-$1906 shows pressure
shifted C$_2$ Swan bands, but the other two with $T_{\rm eff}>6000$ K appear normal. We measure C abundances ranging
from $\log{\rm (C/He)}=-7.9$ to $-3.5$ for these three stars.

\subsection{White Dwarfs with Featureless Optical Spectra or Unknown Spectral Types}

\begin{figure*}
\vspace{-1in}
\includegraphics[angle=-90,width=7in]{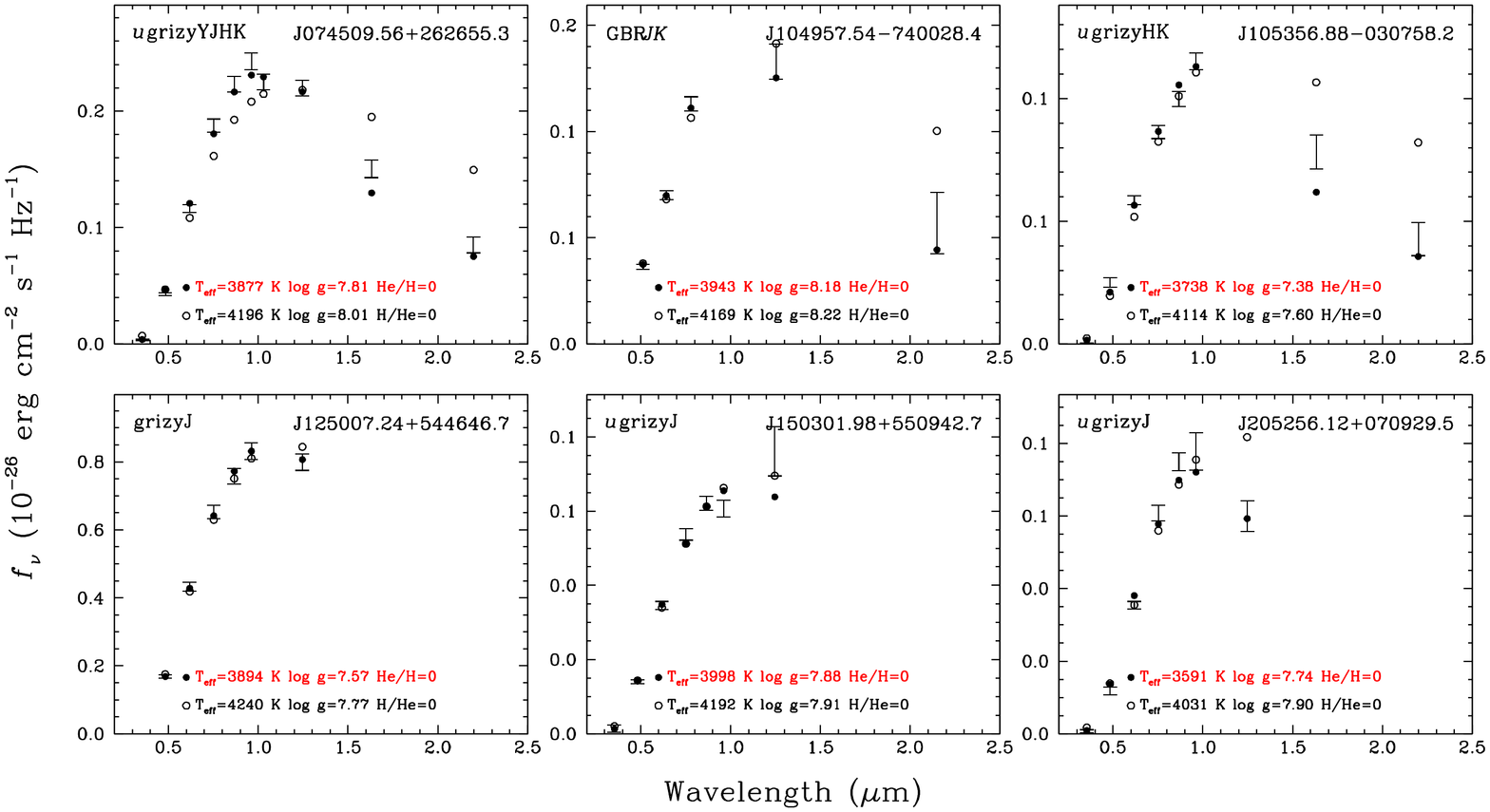}
\vspace{-0.5in}
\caption{SED fits for six ultracool white dwarfs. Only two of these white dwarfs, J0745+2626 and J1250+5446, have optical
spectra available, and both are DC white dwarfs, but given their temperatures we do not expect to see any H absorption lines
anyway. All but one of these objects are clearly H atmosphere white dwarfs based on near-infrared photometry. The SED for
J1503+5509 is not precise enough to favor either of the H or He atmosphere solutions.}
\label{fig:ultra}
\end{figure*}

\ion{He}{I} lines disappear below about $T_{\rm eff}=$ 11 000 K in white dwarf atmospheres, but H lines remain visible for temperatures
as low as 5,000 K \citep{bergeron97}. Hence, DC white dwarfs with effective temperatures between $\approx5000$ K and 11 000 K must have
He-rich atmospheres. Figure \ref{fig:dc} shows our model fits to three DC white dwarfs, including J0248$-$3001 (left panels).
The observed SED for J0248$-$3001 is best-explained by a model with $T_{\rm eff}\approx6000$ K. However, a pure H atmosphere
white dwarf with that temperature would show a significant H$\alpha$ absorption line, which is not observed. Hence,  J0248$-$3001
clearly has a He dominated atmosphere.

The middle and right panels in Figure \ref{fig:dc} demonstrate our choice of composition for two other DC white dwarfs,
WD 0343+247 and WD 0541+260 (J0544+2602). These two white dwarfs are best explained by mixed H/He
atmospheres. J0544+2602 is warm enough to show an H$\alpha$ absorption line if it had a pure H atmosphere. The lack
of H$\alpha$ absorption clearly indicates a He-rich atmosphere, and the observed SED is best explained by
a model that has $\log{(\rm H/He)}=-0.38$. Similarly, the SED for the well known halo white dwarf WD 0343+247
requires a mixed atmosphere with $\log{\rm (H/He)}=0.21$ \citep[see also][]{bergeron01,kilic12}.

The choice of atmospheric composition for white dwarfs cooler than $T_{\rm eff}=5000$ K must rely on optical and near-infrared
photometry, as both pure H and pure He atmosphere white dwarfs show featureless optical spectra at these temperatures.
Collision induced absorption from molecular hydrogen is visible in the near-infrared bands for white dwarfs cooler than about
4000 K \citep{hansen98} for pure H atmospheres, but it becomes visible at hotter temperatures in higher
density environments of He-rich atmospheres. H atmosphere white dwarfs also show significant absorption from the
red-wing of the Ly$\alpha$ line \citep{kowalski06}. Hence, a combination of UV, optical, and near-infrared photometry
can distinguish between H- and He-dominated atmospheres for cool white dwarfs with $T_{\rm eff}\leq5000$ K.

\begin{table*}
\centering
\caption{Best-fit parameters for the halo white dwarf sample. The total ages are estimated based on the pre-white dwarf
evolutionary lifetimes from equation \ref{kal}. Note that this equation over-estimates the total ages for the coolest
white dwarfs.}
\begin{tabular}{cccccccc}
\hline
Name & Spectral & Comp & $T_{\rm eff}$ & Mass & $\log{g}$ & Cooling Age & Total Age \\
     & Type     &      & (K)  & ($M_{\odot}$) & (cm s$^{-2}$) & (Myr) & (Gyr) \\
\hline
J000232.69-321149.1 & DA & H & 20250 $\pm$ 3410 & 0.369$_{-0.050}^{+0.082}$ & 7.420$_{-0.122}^{+0.164}$ & 20$_{-8}^{+49}$ & \dots \\
J000707.96-311337.4 & DB & He & 17470 $\pm$ 3700 & 0.673$_{-0.148}^{+0.229}$ & 8.133$_{-0.184}^{+0.261}$ & 155$_{-93}^{+192}$ & 7.0$_{-3.6}^{+5.8}$ \\
J001311.26-550104.6 & \dots & H & 8870 $\pm$ 300 & 0.455$_{-0.044}^{+0.050}$ & 7.737$_{-0.066}^{+0.070}$ & 598$_{-62}^{+72}$ & \dots \\
J001311.26-550104.6 & \dots & He & 8430 $\pm$ 330 & 0.358$_{-0.041}^{+0.051}$ & 7.569$_{-0.074}^{+0.081}$ & 596$_{-64}^{+76}$ & \dots \\
J003008.73-324402.9 & \dots & H & 9330 $\pm$ 650 & 0.608$_{-0.112}^{+0.134}$ & 8.017$_{-0.143}^{+0.153}$ & 740$_{-172}^{+241}$ & 9.6$_{-3.4}^{+5.6}$ \\
J003008.73-324402.9 & \dots & He & 9150 $\pm$ 640 & 0.543$_{-0.107}^{+0.128}$ & 7.938$_{-0.142}^{+0.150}$ & 723$_{-164}^{+228}$ & 12.5$_{-4.7}^{+7.8}$ \\
J003428.89-684939.3 & \dots & H & 4870 $\pm$ 100 & 0.524$_{-0.039}^{+0.043}$ & 7.907$_{-0.049}^{+0.052}$ & 5211$_{-734}^{+732}$ & 18.0$_{-1.5}^{+1.9}$ \\
J003428.89-684939.3 & \dots & He & 4790 $\pm$ 70 & 0.472$_{-0.034}^{+0.036}$ & 7.832$_{-0.044}^{+0.046}$ & 4616$_{-473}^{+493}$ & \dots \\
J004521.86-332952.0 & \dots & H & 3940 $\pm$ 240 & 0.664$_{-0.090}^{+0.107}$ & 8.137$_{-0.100}^{+0.115}$ & 10100$_{-858}^{+710}$ & 17.1$_{-1.6}^{+2.3}$ \\
J004521.86-332952.0 & \dots & He & 4100 $\pm$ 120 & 0.627$_{-0.077}^{+0.091}$ & 8.095$_{-0.088}^{+0.099}$ & 7844$_{-463}^{+398}$ & 16.0$_{-2.0}^{+2.7}$ \\
J005516.37+384748.2 & DC & He & 5510 $\pm$ 70 & 0.236$_{-0.015}^{+0.016}$ & 7.263$_{-0.037}^{+0.036}$ & 1466$_{-64}^{+68}$ & \dots \\
J010207.56-003301.5 & DA & H & 11220 $\pm$ 430 & 0.623$_{-0.044}^{+0.045}$ & 8.032$_{-0.053}^{+0.054}$ & 466$_{-54}^{+62}$ & 8.8$_{-1.3}^{+1.6}$ \\
J010848.50+151513.1 & DA & H & 18000 $\pm$ 610 & 0.574$_{-0.024}^{+0.025}$ & 7.919$_{-0.032}^{+0.032}$ & 89$_{-13}^{+14}$ & 10.3$_{-1.0}^{+1.1}$ \\
J011745.50+295454.9 & \dots & H & 9480 $\pm$ 250 & 0.355$_{-0.023}^{+0.027}$ & 7.503$_{-0.044}^{+0.047}$ & 410$_{-30}^{+33}$ & \dots \\
J011745.50+295454.9 & \dots & He & 9360 $\pm$ 340 & 0.298$_{-0.023}^{+0.027}$ & 7.395$_{-0.051}^{+0.055}$ & 407$_{-38}^{+43}$ & \dots \\
J013503.59-325954.1 & \dots & H & 13500 $\pm$ 1580 & 0.397$_{-0.052}^{+0.075}$ & 7.570$_{-0.100}^{+0.122}$ & 152$_{-50}^{+78}$ & \dots \\
J013503.59-325954.1 & \dots & He & 13390 $\pm$ 1420 & 0.372$_{-0.059}^{+0.092}$ & 7.564$_{-0.117}^{+0.147}$ & 164$_{-48}^{+73}$ & \dots \\
J013534.20-035721.1 & DA & H & 6640 $\pm$ 130 & 0.658$_{-0.075}^{+0.073}$ & 8.110$_{-0.087}^{+0.081}$ & 2136$_{-363}^{+458}$ & 9.4$_{-1.3}^{+2.3}$ \\
J014220.89-012356.6 & \dots & H & 9880 $\pm$ 270 & 0.710$_{-0.076}^{+0.073}$ & 8.178$_{-0.087}^{+0.083}$ & 807$_{-112}^{+123}$ & 6.7$_{-1.2}^{+1.9}$ \\
J014809.04-171231.6 & DA & H & 7190 $\pm$ 140 & 0.498$_{-0.029}^{+0.031}$ & 7.833$_{-0.039}^{+0.041}$ & 1124$_{-77}^{+87}$ & \dots \\
J015351.57-012347.6 & DA & H & 9230 $\pm$ 230 & 0.663$_{-0.064}^{+0.063}$ & 8.107$_{-0.076}^{+0.072}$ & 869$_{-107}^{+113}$ & 7.9$_{-1.4}^{+2.0}$ \\
J020306.58+354008.1 & \dots & H & 7400 $\pm$ 220 & 0.544$_{-0.038}^{+0.043}$ & 7.917$_{-0.050}^{+0.052}$ & 1174$_{-117}^{+131}$ & 12.9$_{-1.9}^{+2.1}$ \\
J020306.58+354008.1 & \dots & He & 7050 $\pm$ 210 & 0.445$_{-0.039}^{+0.044}$ & 7.767$_{-0.056}^{+0.059}$ & 1140$_{-111}^{+127}$ & \dots\\
J020337.73-045915.7 & DA & H & 24720 $\pm$ 930 & 0.484$_{-0.021}^{+0.024}$ & 7.695$_{-0.036}^{+0.039}$ & 17$_{-2}^{+2}$ & \dots \\
J023438.80-594417.0 & \dots & H & 8580 $\pm$ 430 & 0.491$_{-0.080}^{+0.095}$ & 7.811$_{-0.116}^{+0.122}$ & 704$_{-114}^{+147}$ & \dots \\
J023438.80-594417.0 & \dots & He & 8280 $\pm$ 450 & 0.412$_{-0.076}^{+0.097}$ & 7.693$_{-0.126}^{+0.133}$ & 698$_{-115}^{+156}$ & \dots \\
J023737.80-844521.4 & \dots & [H/He=1.84] & 3840 $\pm$ 170 & 0.425$_{-0.066}^{+0.080}$ & 7.745$_{-0.098}^{+0.106}$ & 6395$_{-1027}^{+860}$ & \dots \\
J024138.84-482144.9 & \dots & H & 13910 $\pm$ 1420 & 0.570$_{-0.095}^{+0.110}$ & 7.930$_{-0.131}^{+0.135}$ & 223$_{-71}^{+96}$ & 10.6$_{-3.7}^{+5.7}$ \\
J024138.84-482144.9 & \dots & He & 13380 $\pm$ 1240 & 0.524$_{-0.101}^{+0.125}$ & 7.888$_{-0.145}^{+0.154}$ & 239$_{-72}^{+100}$ & 13.0$_{-5.2}^{+8.2}$ \\
J024813.73-300127.8 & DC & He & 6020 $\pm$ 160 & 0.521$_{-0.043}^{+0.048}$ & 7.913$_{-0.054}^{+0.058}$ & 2155$_{-269}^{+350}$ & 15.1$_{-2.2}^{+2.7}$ \\
J025754.76+075101.6 & \dots & H & 8990 $\pm$ 350 & 0.580$_{-0.079}^{+0.085}$ & 7.971$_{-0.102}^{+0.100}$ & 763$_{-116}^{+145}$ & 10.7$_{-2.8}^{+4.2}$ \\
J025754.76+075101.6 & \dots & He & 8820 $\pm$ 400 & 0.514$_{-0.080}^{+0.089}$ & 7.889$_{-0.110}^{+0.109}$ & 750$_{-125}^{+149}$ & 14.2$_{-4.2}^{+6.4}$ \\
J030023.97-042529.8 & DC & He & 5860 $\pm$ 110 & 0.526$_{-0.072}^{+0.074}$ & 7.924$_{-0.094}^{+0.088}$ & 2446$_{-424}^{+568}$ & 15.1$_{-3.0}^{+4.9}$ \\
J030144.18-004446.6 & DC & H & 4340 $\pm$ 80 & 0.451$_{-0.045}^{+0.049}$ & 7.780$_{-0.063}^{+0.063}$ & 5766$_{-722}^{+758}$ & \dots \\
J031000.79+163018.0 & DA & H & 17960 $\pm$ 2010 & 0.482$_{-0.053}^{+0.069}$ & 7.735$_{-0.086}^{+0.100}$ & 64$_{-23}^{+41}$ & \dots \\
J032228.50+835224.1 & \dots & H & 6170 $\pm$ 180 & 0.573$_{-0.053}^{+0.057}$ & 7.973$_{-0.065}^{+0.068}$ & 1989$_{-224}^{+324}$ & 12.3$_{-1.9}^{+2.5}$ \\
J032228.50+835224.1 & \dots & He & 6030 $\pm$ 170 & 0.504$_{-0.052}^{+0.059}$ & 7.884$_{-0.068}^{+0.072}$ & 2039$_{-267}^{+353}$ & 16.1$_{-3.0}^{+3.8}$ \\
J034200.81+361556.0 & \dots & H & 19580 $\pm$ 830 & 0.506$_{-0.037}^{+0.042}$ & 7.777$_{-0.059}^{+0.060}$ & 46$_{-8}^{+11}$ & 14.0$_{-2.4}^{+2.7}$ \\
J034532.94-361113.1 & DC & He & 4300 $\pm$ 60 & 0.420$_{-0.042}^{+0.047}$ & 7.736$_{-0.063}^{+0.063}$ & 5006$_{-536}^{+551}$ & \dots \\
J034647.11+245544.7 & DC & [H/He=0.21] & 3580 $\pm$ 40 & 0.331$_{-0.020}^{+0.022}$ & 7.545$_{-0.035}^{+0.036}$ & 5364$_{-331}^{+357}$ & \dots \\
J034856.92-034701.4 & \dots & H & 6410 $\pm$ 180 & 0.630$_{-0.070}^{+0.071}$ & 8.066$_{-0.083}^{+0.080}$ & 2165$_{-367}^{+490}$ & 10.2$_{-1.5}^{+2.4}$ \\
J034856.92-034701.4 & \dots & He & 6220 $\pm$ 170 & 0.548$_{-0.070}^{+0.073}$ & 7.959$_{-0.089}^{+0.085}$ & 2036$_{-283}^{+412}$ & 13.5$_{-2.7}^{+4.2}$ \\
J035228.87-412657.8 & \dots & H & 5700 $\pm$ 150 & 0.610$_{-0.076}^{+0.078}$ & 8.039$_{-0.091}^{+0.089}$ & 2898$_{-568}^{+867}$ & 11.7$_{-1.5}^{+2.9}$ \\
J035228.87-412657.8 & \dots & He & 5460 $\pm$ 140 & 0.501$_{-0.074}^{+0.078}$ & 7.882$_{-0.097}^{+0.095}$ & 3114$_{-598}^{+862}$ & 17.4$_{-3.4}^{+5.8}$ \\
J041309.57-571341.1 & \dots & H & 10940 $\pm$ 640 & 0.521$_{-0.064}^{+0.075}$ & 7.854$_{-0.090}^{+0.096}$ & 398$_{-69}^{+84}$ & 13.4$_{-3.6}^{+4.7}$ \\
J041309.57-571341.1 & \dots & He & 10810 $\pm$ 680 & 0.447$_{-0.064}^{+0.080}$ & 7.750$_{-0.100}^{+0.108}$ & 370$_{-68}^{+88}$ & \dots \\
J043236.94-390203.0 & DC & H & 4580 $\pm$ 100 & 0.404$_{-0.038}^{+0.044}$ & 7.685$_{-0.057}^{+0.061}$ & 3893$_{-465}^{+668}$ & \dots \\
J045400.36-432246.6 & \dots & H & 11120 $\pm$ 450 & 0.417$_{-0.025}^{+0.028}$ & 7.637$_{-0.040}^{+0.044}$ & 296$_{-34}^{+39}$ & \dots \\
J045400.36-432246.6 & \dots & He & 10940 $\pm$ 460 & 0.347$_{-0.025}^{+0.031}$ & 7.518$_{-0.050}^{+0.055}$ & 290$_{-33}^{+38}$ & \dots \\
J045823.12-563733.5 & \dots & H & 26760 $\pm$ 2890 & 0.394$_{-0.041}^{+0.064}$ & 7.427$_{-0.116}^{+0.136}$ & 14$_{-4}^{+5}$ & \dots \\
J045823.12-563733.5 & \dots & He & 32390 $\pm$ 4770 & 0.415$_{-0.049}^{+0.075}$ & 7.530$_{-0.120}^{+0.144}$ & 9$_{-5}^{+5}$ & \dots \\
J053712.48+321501.0 & DA & H & 11580 $\pm$ 590 & 0.236$_{-0.010}^{+0.011}$ & 7.041$_{-0.037}^{+0.039}$ & 155$_{-23}^{+27}$ & \dots \\
J054458.81+260238.6 & DC & [H/He=$-$0.38] & 4830 $\pm$ 70 & 0.238$_{-0.017}^{+0.019}$ & 7.281$_{-0.041}^{+0.041}$ & 2090$_{-95}^{+98}$ & \dots \\
J054517.81-414823.3 & \dots & H & 6450 $\pm$ 240 & 0.512$_{-0.080}^{+0.090}$ & 7.864$_{-0.109}^{+0.111}$ & 1534$_{-235}^{+293}$ & 15.1$_{-4.2}^{+6.4}$ \\
J054517.81-414823.3 & \dots & He & 6220 $\pm$ 220 & 0.424$_{-0.077}^{+0.088}$ & 7.729$_{-0.118}^{+0.119}$ & 1500$_{-234}^{+311}$ & \dots \\
J054936.61+232938.9 & \dots & He & 4730 $\pm$ 40 & 0.306$_{-0.015}^{+0.015}$ & 7.476$_{-0.027}^{+0.028}$ & 2605$_{-86}^{+93}$ & \dots \\
J055038.28-361633.8 & \dots & H & 20260 $\pm$ 1760 & 0.609$_{-0.049}^{+0.058}$ & 7.973$_{-0.065}^{+0.073}$ & 61$_{-23}^{+32}$ & 8.9$_{-1.8}^{+2.0}$ \\
J055038.28-361633.8 & \dots & He & 19980 $\pm$ 2700 & 0.630$_{-0.072}^{+0.091}$ & 8.055$_{-0.091}^{+0.108}$ & 79$_{-39}^{+63}$ & 8.2$_{-2.3}^{+2.9}$ \\
J055910.47+424838.1 & \dots & H & 6870 $\pm$ 190 & 0.563$_{-0.048}^{+0.052}$ & 7.952$_{-0.061}^{+0.062}$ & 1483$_{-153}^{+177}$ & 12.2$_{-2.0}^{+2.5}$ \\
J055910.47+424838.1 & \dots & He & 6520 $\pm$ 180 & 0.452$_{-0.046}^{+0.052}$ & 7.784$_{-0.066}^{+0.068}$ & 1422$_{-151}^{+167}$ & \dots \\
\hline
\end{tabular}
\end{table*}

\begin{table*}
\centering
\contcaption{}
\begin{tabular}{cccccccc}
\hline
Name & Spectral & Comp & $T_{\rm eff}$ & Mass & $\log{g}$ & Cooling Age & Total Age \\
     & Type     &      & (K)  & ($M_{\odot}$) & (cm s$^{-2}$) & (Myr) & (Gyr) \\
\hline
J061105.02-252027.9 & \dots & H & 7000 $\pm$ 250 & 0.631$_{-0.074}^{+0.077}$ & 8.064$_{-0.088}^{+0.088}$ & 1665$_{-246}^{+325}$ & 9.7$_{-1.8}^{+2.7}$ \\
J061105.02-252027.9 & \dots & He & 6800 $\pm$ 240 & 0.555$_{-0.075}^{+0.082}$ & 7.969$_{-0.095}^{+0.094}$ & 1595$_{-211}^{+309}$ & 12.7$_{-3.0}^{+4.5}$ \\
J064832.45+302639.1 & \dots & H & 9740 $\pm$ 370 & 0.449$_{-0.070}^{+0.079}$ & 7.720$_{-0.112}^{+0.110}$ & 463$_{-65}^{+78}$ & \dots \\
J064832.45+302639.1 & \dots & He & 9810 $\pm$ 490 & 0.399$_{-0.069}^{+0.081}$ & 7.653$_{-0.120}^{+0.119}$ & 434$_{-66}^{+86}$ & \dots \\
J071147.80+460734.7 & DC & He & 4820 $\pm$ 70 & 0.435$_{-0.037}^{+0.040}$ & 7.763$_{-0.053}^{+0.054}$ & 3872$_{-458}^{+548}$ & \dots \\
J072112.45+104219.8 & \dots & H & 13260 $\pm$ 1060 & 0.389$_{-0.054}^{+0.067}$ & 7.552$_{-0.107}^{+0.111}$ & 158$_{-38}^{+52}$ & \dots \\
J072112.45+104219.8 & \dots & He & 14460 $\pm$ 1220 & 0.420$_{-0.070}^{+0.093}$ & 7.673$_{-0.125}^{+0.135}$ & 140$_{-38}^{+54}$ & \dots \\
J074509.56+262655.3 & DC & H & 3880 $\pm$ 60 & 0.468$_{-0.030}^{+0.032}$ & 7.814$_{-0.040}^{+0.041}$ & 7707$_{-435}^{+427}$ & \dots \\
J074521.02+685653.9 & \dots & H & 5720 $\pm$ 140 & 0.578$_{-0.049}^{+0.054}$ & 7.987$_{-0.060}^{+0.062}$ & 2545$_{-332}^{+486}$ & 12.6$_{-1.6}^{+2.1}$ \\
J074521.02+685653.9 & \dots & He & 5600 $\pm$ 130 & 0.510$_{-0.048}^{+0.053}$ & 7.898$_{-0.062}^{+0.064}$ & 2841$_{-417}^{+603}$ & 16.5$_{-2.3}^{+3.2}$ \\
J075014.80+071121.3 & DC & He & 4430 $\pm$ 50 & 0.479$_{-0.026}^{+0.028}$ & 7.847$_{-0.034}^{+0.035}$ & 5584$_{-321}^{+326}$ & \dots \\
J075015.56+071109.3 & DC & He & 4730 $\pm$ 60 & 0.524$_{-0.026}^{+0.028}$ & 7.926$_{-0.033}^{+0.034}$ & 5643$_{-313}^{+296}$ & 18.4$_{-1.2}^{+1.4}$ \\
J082219.34-124924.0 & \dots & H & 4120 $\pm$ 160 & 0.453$_{-0.058}^{+0.069}$ & 7.785$_{-0.082}^{+0.088}$ & 6583$_{-999}^{+1042}$ & \dots \\
J082219.34-124924.0 & \dots & He & 4260 $\pm$ 90 & 0.459$_{-0.046}^{+0.051}$ & 7.809$_{-0.063}^{+0.066}$ & 5720$_{-564}^{+565}$ & \dots \\
J084054.32+145706.6 & DA & H & 11070 $\pm$ 390 & 0.582$_{-0.067}^{+0.069}$ & 7.963$_{-0.087}^{+0.083}$ & 441$_{-58}^{+67}$ & 10.3$_{-2.4}^{+3.4}$ \\
J084623.66+492532.0 & DA & H & 9050 $\pm$ 260 & 0.340$_{-0.045}^{+0.053}$ & 7.468$_{-0.094}^{+0.094}$ & 451$_{-43}^{+48}$ & \dots \\
J091304.64-375131.0 & \dots & H & 9420 $\pm$ 540 & 0.571$_{-0.074}^{+0.090}$ & 7.953$_{-0.096}^{+0.107}$ & 661$_{-114}^{+150}$ & 11.0$_{-3.1}^{+4.1}$ \\
J091304.64-375131.0 & \dots & He & 9250 $\pm$ 570 & 0.504$_{-0.072}^{+0.089}$ & 7.868$_{-0.099}^{+0.111}$ & 644$_{-119}^{+151}$ & 14.7$_{-4.5}^{+6.0}$ \\
J091523.36-214923.6 & \dots & H & 5850 $\pm$ 110 & 0.494$_{-0.032}^{+0.035}$ & 7.837$_{-0.043}^{+0.045}$ & 1899$_{-139}^{+149}$ & \dots \\
J091523.36-214923.6 & \dots & He & 5620 $\pm$ 100 & 0.395$_{-0.030}^{+0.033}$ & 7.677$_{-0.046}^{+0.047}$ & 2069$_{-183}^{+191}$ & \dots \\
J092531.22+001814.7 & \dots & [H/He=0.07] & 4220 $\pm$ 70 & 0.360$_{-0.035}^{+0.038}$ & 7.609$_{-0.058}^{+0.059}$ & 4199$_{-452}^{+499}$ & \dots \\
J093900.89+391609.4 & \dots & H & 4910 $\pm$ 120 & 0.382$_{-0.080}^{+0.087}$ & 7.633$_{-0.135}^{+0.124}$ & 2789$_{-533}^{+802}$ & \dots \\
J093900.89+391609.4 & \dots & He & 4880 $\pm$ 90 & 0.361$_{-0.076}^{+0.084}$ & 7.609$_{-0.135}^{+0.122}$ & 2789$_{-393}^{+708}$ & \dots \\
J094129.72+651131.9 & \dots & H & 4400 $\pm$ 90 & 0.427$_{-0.034}^{+0.038}$ & 7.733$_{-0.049}^{+0.051}$ & 5022$_{-578}^{+643}$ & \dots \\
J094129.72+651131.9 & \dots & He & 4470 $\pm$ 60 & 0.429$_{-0.032}^{+0.035}$ & 7.752$_{-0.046}^{+0.047}$ & 4670$_{-406}^{+444}$ & \dots \\
J095651.56+250546.3 & \dots & H & 5970 $\pm$ 260 & 0.439$_{-0.092}^{+0.106}$ & 7.729$_{-0.144}^{+0.140}$ & 1565$_{-280}^{+354}$ & \dots \\
J095651.56+250546.3 & \dots & He & 5790 $\pm$ 240 & 0.366$_{-0.085}^{+0.103}$ & 7.612$_{-0.154}^{+0.148}$ & 1694$_{-348}^{+433}$ & \dots \\
J100514.02+025416.9 & \dots & H & 4510 $\pm$ 80 & 0.513$_{-0.087}^{+0.087}$ & 7.891$_{-0.114}^{+0.103}$ & 6444$_{-1388}^{+1090}$ & 19.9$_{-3.2}^{+5.9}$ \\
J100514.02+025416.9 & \dots & He & 4550 $\pm$ 60 & 0.505$_{-0.085}^{+0.085}$ & 7.894$_{-0.113}^{+0.100}$ & 5718$_{-1038}^{+798}$ & 19.7$_{-3.7}^{+6.4}$ \\
J100817.57-430633.1 & \dots & H & 9760 $\pm$ 530 & 0.526$_{-0.063}^{+0.078}$ & 7.870$_{-0.087}^{+0.098}$ & 545$_{-87}^{+109}$ & 13.2$_{-3.6}^{+4.4}$ \\
J100817.57-430633.1 & \dots & He & 9480 $\pm$ 570 & 0.442$_{-0.059}^{+0.075}$ & 7.747$_{-0.091}^{+0.102}$ & 523$_{-89}^{+114}$ & \dots \\
J103654.94+073210.8 & DC & H & 4260 $\pm$ 90 & 0.388$_{-0.038}^{+0.044}$ & 7.655$_{-0.060}^{+0.062}$ & 4803$_{-593}^{+657}$ & \dots \\
J103654.94+073210.8 & DC & He & 4390 $\pm$ 60 & 0.409$_{-0.038}^{+0.042}$ & 7.712$_{-0.057}^{+0.058}$ & 4554$_{-463}^{+509}$ & \dots \\
J103920.88+054358.9 & DC & He & 7050 $\pm$ 110 & 0.649$_{-0.065}^{+0.063}$ & 8.121$_{-0.074}^{+0.069}$ & 1892$_{-272}^{+333}$ & 9.4$_{-1.3}^{+2.0}$ \\
J104006.88+244227.9 & \dots & H & 6540 $\pm$ 140 & 0.419$_{-0.074}^{+0.080}$ & 7.681$_{-0.121}^{+0.112}$ & 1191$_{-145}^{+180}$ & \dots \\
J104006.88+244227.9 & \dots & He & 6460 $\pm$ 140 & 0.374$_{-0.074}^{+0.080}$ & 7.620$_{-0.128}^{+0.119}$ & 1193$_{-140}^{+190}$ & \dots \\
J104537.00-190654.4 & DQ & [C/He=$-$7.92] & 5270 $\pm$ 60 & 0.327$_{-0.014}^{+0.015}$ & 7.526$_{-0.025}^{+0.025}$ & 2110$_{-80}^{+85}$ & \dots \\
J104557.37+590428.6 & DQ & [C/He=$-$3.51] & 8690 $\pm$ 120 & 1.029$_{-0.015}^{+0.016}$ & 8.701$_{-0.020}^{+0.020}$ & 2715$_{-80}^{+81}$ & 4.8$_{-0.0}^{+0.0}$ \\
J104957.54-740028.4 & \dots & H & 3940 $\pm$ 180 & 0.694$_{-0.069}^{+0.079}$ & 8.183$_{-0.076}^{+0.085}$ & 10310$_{-558}^{+480}$ & 16.6$_{-1.1}^{+1.4}$ \\
J105356.88-030758.2 & \dots & H & 3740 $\pm$ 100 & 0.272$_{-0.039}^{+0.045}$ & 7.377$_{-0.085}^{+0.084}$ & 3897$_{-408}^{+615}$ & \dots \\
J110114.85+633345.8 & \dots & H & 4430 $\pm$ 100 & 0.506$_{-0.054}^{+0.058}$ & 7.879$_{-0.069}^{+0.070}$ & 6563$_{-870}^{+808}$ & 20.5$_{-2.4}^{+3.4}$ \\
J110114.85+633345.8 & \dots & He & 4480 $\pm$ 70 & 0.495$_{-0.051}^{+0.056}$ & 7.876$_{-0.067}^{+0.068}$ & 5736$_{-619}^{+572}$ & \dots \\
J110704.77+040907.5 & DA & H & 22860 $\pm$ 910 & 0.570$_{-0.031}^{+0.033}$ & 7.887$_{-0.044}^{+0.045}$ & 27$_{-5}^{+7}$ & 10.4$_{-1.4}^{+1.5}$ \\
J111248.13-750035.4 & \dots & H & 11240 $\pm$ 660 & 0.296$_{-0.020}^{+0.026}$ & 7.303$_{-0.054}^{+0.058}$ & 222$_{-35}^{+42}$ & \dots \\
J111248.13-750035.4 & \dots & He & 11200 $\pm$ 600 & 0.255$_{-0.021}^{+0.027}$ & 7.217$_{-0.063}^{+0.072}$ & 222$_{-33}^{+39}$ & \dots \\
J114730.20-745738.2 & \dots & H & 4210 $\pm$ 80 & 0.600$_{-0.038}^{+0.041}$ & 8.038$_{-0.044}^{+0.046}$ & 8781$_{-436}^{+401}$ & 17.9$_{-1.0}^{+1.2}$ \\
J115131.13+015952.6 & \dots & H & 4540 $\pm$ 90 & 0.474$_{-0.061}^{+0.065}$ & 7.821$_{-0.083}^{+0.081}$ & 5499$_{-1004}^{+1022}$ & \dots \\
J115131.13+015952.6 & \dots & He & 4580 $\pm$ 70 & 0.469$_{-0.059}^{+0.063}$ & 7.829$_{-0.081}^{+0.079}$ & 5069$_{-756}^{+742}$ & \dots \\
J115941.74-463034.3 & \dots & H & 4610 $\pm$ 140 & 0.568$_{-0.065}^{+0.075}$ & 7.984$_{-0.078}^{+0.085}$ & 7067$_{-954}^{+911}$ & 17.6$_{-1.9}^{+2.7}$ \\
J115941.74-463034.3 & \dots & He & 4600 $\pm$ 100 & 0.542$_{-0.057}^{+0.064}$ & 7.956$_{-0.069}^{+0.074}$ & 6129$_{-615}^{+554}$ & 17.9$_{-2.3}^{+3.0}$ \\
J115956.86-462903.3 & DQ & [C/He=$-$6.17] & 6780 $\pm$ 170 & 0.533$_{-0.035}^{+0.038}$ & 7.930$_{-0.044}^{+0.046}$ & 1530$_{-117}^{+135}$ & 13.8$_{-1.8}^{+2.1}$ \\
J124023.32-231756.1 & DA & H & 5530 $\pm$ 40 & 0.202$_{-0.007}^{+0.008}$ & 7.082$_{-0.021}^{+0.021}$ & 1125$_{-23}^{+24}$ & \dots \\
J125007.24+544646.7 & DC & H & 3890 $\pm$ 90 & 0.351$_{-0.032}^{+0.036}$ & 7.575$_{-0.054}^{+0.057}$ & 5091$_{-586}^{+685}$ & \dots \\
J125506.81+465517.1 & \dots & H & 4810 $\pm$ 60 & 0.802$_{-0.025}^{+0.027}$ & 8.342$_{-0.028}^{+0.028}$ & 8913$_{-182}^{+171}$ & 13.2$_{-0.2}^{+0.2}$ \\
J125506.81+465517.1 & \dots & He & 4630 $\pm$ 60 & 0.702$_{-0.027}^{+0.028}$ & 8.209$_{-0.029}^{+0.030}$ & 7454$_{-129}^{+115}$ & 13.5$_{-0.5}^{+0.5}$ \\
J125816.72+000708.0 & DA & H & 15250 $\pm$ 560 & 0.566$_{-0.051}^{+0.052}$ & 7.915$_{-0.068}^{+0.067}$ & 163$_{-26}^{+28}$ & 10.8$_{-2.1}^{+2.7}$ \\
J131253.16-472808.9 & DC & H & 4240 $\pm$ 80 & 0.558$_{-0.039}^{+0.043}$ & 7.969$_{-0.046}^{+0.049}$ & 8096$_{-522}^{+505}$ & 19.1$_{-1.4}^{+1.6}$ \\
J131643.36-153608.7 & DA & H & 15210 $\pm$ 880 & 0.543$_{-0.028}^{+0.031}$ & 7.873$_{-0.038}^{+0.041}$ & 153$_{-29}^{+35}$ & 11.9$_{-1.5}^{+1.6}$ \\
J132019.90-200652.6 & \dots & He & 4720 $\pm$ 60 & 0.400$_{-0.083}^{+0.086}$ & 7.693$_{-0.134}^{+0.117}$ & 3510$_{-667}^{+1034}$ & \dots \\
J132325.02+303607.3 & DA & H & 6670 $\pm$ 120 & 0.490$_{-0.056}^{+0.058}$ & 7.822$_{-0.078}^{+0.075}$ & 1338$_{-136}^{+149}$ & \dots \\
J133232.96-241118.1 & \dots & H & 7430 $\pm$ 200 & 0.336$_{-0.040}^{+0.046}$ & 7.476$_{-0.080}^{+0.080}$ & 736$_{-62}^{+69}$ & \dots \\
J133232.96-241118.1 & \dots & He & 7100 $\pm$ 200 & 0.262$_{-0.035}^{+0.041}$ & 7.317$_{-0.086}^{+0.086}$ & 769$_{-59}^{+66}$ & \dots \\
\hline
\end{tabular}
\end{table*}

\begin{table*}
\centering
\contcaption{}
\begin{tabular}{cccccccc}
\hline
Name & Spectral & Comp & $T_{\rm eff}$ & Mass & $\log{g}$ & Cooling Age & Total Age \\
     & Type     &      & (K)  & ($M_{\odot}$) & (cm s$^{-2}$) & (Myr) & (Gyr) \\
\hline
J134160.00-341501.7 & DA & H & 5210 $\pm$ 90 & 0.598$_{-0.029}^{+0.031}$ & 8.027$_{-0.034}^{+0.036}$ & 4962$_{-657}^{+624}$ & 14.2$_{-0.5}^{+0.6}$ \\
J134551.88+400100.9 & DA & H & 17590 $\pm$ 730 & 0.558$_{-0.069}^{+0.071}$ & 7.890$_{-0.098}^{+0.092}$ & 92$_{-21}^{+25}$ & 11.1$_{-2.8}^{+4.1}$ \\
J141353.86+153020.1 & DA & H & 14310 $\pm$ 460 & 0.319$_{-0.012}^{+0.014}$ & 7.337$_{-0.029}^{+0.032}$ & 108$_{-11}^{+12}$ & \dots \\
J141542.83+225644.9 & DA & H & 24400 $\pm$ 1020 & 0.547$_{-0.027}^{+0.029}$ & 7.833$_{-0.039}^{+0.042}$ & 19$_{-3}^{+4}$ & 11.5$_{-1.4}^{+1.5}$ \\
J142433.94+190634.0 & DA & H & 15610 $\pm$ 570 & 0.517$_{-0.039}^{+0.041}$ & 7.821$_{-0.056}^{+0.057}$ & 129$_{-20}^{+22}$ & 13.4$_{-2.2}^{+2.7}$ \\
J142649.92+094932.6 & \dots & H & 15970 $\pm$ 540 & 0.505$_{-0.047}^{+0.050}$ & 7.796$_{-0.072}^{+0.069}$ & 113$_{-18}^{+21}$ & 14.1$_{-2.9}^{+3.6}$ \\
J144101.58+013210.3 & \dots & H & 10900 $\pm$ 360 & 0.503$_{-0.097}^{+0.101}$ & 7.819$_{-0.145}^{+0.130}$ & 386$_{-69}^{+74}$ & 14.5$_{-5.1}^{+8.9}$ \\
J144205.74+220323.9 & \dots & H & 8130 $\pm$ 320 & 0.414$_{-0.085}^{+0.098}$ & 7.657$_{-0.146}^{+0.140}$ & 688$_{-100}^{+125}$ & \dots \\
J144205.74+220323.9 & \dots & He & 7900 $\pm$ 250 & 0.350$_{-0.075}^{+0.091}$ & 7.555$_{-0.149}^{+0.140}$ & 693$_{-85}^{+111}$ & \dots \\
J144350.27+505934.5 & \dots & H & 6650 $\pm$ 140 & 0.578$_{-0.051}^{+0.054}$ & 7.979$_{-0.064}^{+0.064}$ & 1668$_{-169}^{+209}$ & 11.7$_{-1.8}^{+2.4}$ \\
J144350.27+505934.5 & \dots & He & 6550 $\pm$ 150 & 0.521$_{-0.053}^{+0.057}$ & 7.911$_{-0.068}^{+0.068}$ & 1638$_{-152}^{+172}$ & 14.6$_{-2.8}^{+3.6}$ \\
J145048.53+073326.9 & DA & H & 15120 $\pm$ 440 & 0.553$_{-0.017}^{+0.018}$ & 7.892$_{-0.023}^{+0.023}$ & 161$_{-16}^{+17}$ & 11.4$_{-0.8}^{+0.9}$ \\
J145738.18+242336.7 & \dots & H & 13320 $\pm$ 530 & 0.554$_{-0.058}^{+0.058}$ & 7.902$_{-0.078}^{+0.075}$ & 245$_{-37}^{+41}$ & 11.4$_{-2.4}^{+3.4}$ \\
J150002.23+360017.9 & DC & H & 4860 $\pm$ 60 & 0.411$_{-0.028}^{+0.031}$ & 7.696$_{-0.042}^{+0.044}$ & 3258$_{-306}^{+356}$ & \dots \\
J150002.23+360017.9 & DC & He & 4820 $\pm$ 50 & 0.383$_{-0.026}^{+0.028}$ & 7.657$_{-0.041}^{+0.041}$ & 3089$_{-204}^{+260}$ & \dots \\
J150301.98+550942.7 & \dots & H & 4000 $\pm$ 120 & 0.505$_{-0.056}^{+0.061}$ & 7.879$_{-0.071}^{+0.074}$ & 7903$_{-772}^{+795}$ & 21.9$_{-2.6}^{+3.7}$ \\
J150301.98+550942.7 & \dots & He & 4190 $\pm$ 70 & 0.515$_{-0.050}^{+0.053}$ & 7.911$_{-0.064}^{+0.064}$ & 6653$_{-478}^{+439}$ & 20.0$_{-2.4}^{+3.2}$ \\
J150629.55-402631.0 & \dots & H & 4590 $\pm$ 150 & 0.511$_{-0.123}^{+0.132}$ & 7.887$_{-0.166}^{+0.154}$ & 6115$_{-2014}^{+1623}$ & 19.7$_{-4.3}^{+9.9}$ \\
J150629.55-402631.0 & \dots & He & 4610 $\pm$ 120 & 0.499$_{-0.121}^{+0.128}$ & 7.882$_{-0.166}^{+0.150}$ & 5494$_{-1529}^{+1181}$ & \dots \\
J151320.99+474322.5 & \dots & H & 5910 $\pm$ 100 & 0.519$_{-0.051}^{+0.054}$ & 7.883$_{-0.068}^{+0.066}$ & 1958$_{-178}^{+202}$ & 15.1$_{-2.6}^{+3.5}$ \\
J151320.99+474322.5 & \dots & He & 5780 $\pm$ 100 & 0.452$_{-0.053}^{+0.055}$ & 7.788$_{-0.074}^{+0.072}$ & 2069$_{-216}^{+279}$ & \dots \\
J151530.80+191121.2 & \dots & He & 4620 $\pm$ 60 & 0.420$_{-0.046}^{+0.049}$ & 7.734$_{-0.068}^{+0.066}$ & 4106$_{-522}^{+614}$ & \dots \\
J153034.17-750526.2 & DA & H & 21960 $\pm$ 1970 & 0.493$_{-0.041}^{+0.051}$ & 7.734$_{-0.067}^{+0.077}$ & 27$_{-8}^{+15}$ & \dots \\
J153719.34+223720.8 & \dots & H & 4370 $\pm$ 100 & 0.297$_{-0.089}^{+0.103}$ & 7.441$_{-0.195}^{+0.169}$ & 3147$_{-597}^{+942}$ & \dots \\
J153719.34+223720.8 & \dots & He & 4530 $\pm$ 70 & 0.341$_{-0.096}^{+0.108}$ & 7.565$_{-0.185}^{+0.160}$ & 3249$_{-671}^{+1025}$ & \dots \\
J160114.77+534609.2 & DA & H & 7200 $\pm$ 130 & 0.521$_{-0.029}^{+0.030}$ & 7.875$_{-0.037}^{+0.038}$ & 1184$_{-79}^{+87}$ & 14.2$_{-1.6}^{+1.8}$ \\
J163731.64+010031.5 & \dots & H & 6330 $\pm$ 120 & 0.514$_{-0.065}^{+0.068}$ & 7.870$_{-0.088}^{+0.084}$ & 1622$_{-183}^{+205}$ & 15.0$_{-3.3}^{+4.8}$ \\
J163731.64+010031.5 & \dots & He & 6240 $\pm$ 120 & 0.463$_{-0.066}^{+0.069}$ & 7.806$_{-0.094}^{+0.088}$ & 1645$_{-211}^{+219}$ & \dots \\
J164413.61+270120.1 & DA & H & 24400 $\pm$ 2940 & 0.537$_{-0.059}^{+0.081}$ & 7.814$_{-0.094}^{+0.113}$ & 19$_{-7}^{+14}$ & 12.1$_{-3.6}^{+3.9}$ \\
J165705.14+860044.3 & \dots & H & 7360 $\pm$ 300 & 0.688$_{-0.065}^{+0.070}$ & 8.153$_{-0.074}^{+0.079}$ & 1663$_{-239}^{+396}$ & 8.1$_{-1.1}^{+1.7}$ \\
J165705.14+860044.3 & \dots & He & 7180 $\pm$ 290 & 0.619$_{-0.067}^{+0.074}$ & 8.072$_{-0.079}^{+0.083}$ & 1631$_{-251}^{+348}$ & 10.1$_{-1.8}^{+2.6}$ \\
J165708.82+205605.6 & DA & H & 34140 $\pm$ 2690 & 0.452$_{-0.030}^{+0.034}$ & 7.528$_{-0.065}^{+0.073}$ & 7$_{-1}^{+2}$ & \dots \\
J171315.03+520618.7 & \dots & H & 6010 $\pm$ 190 & 0.488$_{-0.096}^{+0.104}$ & 7.824$_{-0.138}^{+0.130}$ & 1744$_{-310}^{+351}$ & \dots \\
J171315.03+520618.7 & \dots & He & 5890 $\pm$ 190 & 0.429$_{-0.095}^{+0.105}$ & 7.744$_{-0.148}^{+0.137}$ & 1815$_{-326}^{+469}$ & \dots \\
J171529.75-732353.0 & \dots & H & 4950 $\pm$ 170 & 0.806$_{-0.084}^{+0.089}$ & 8.347$_{-0.091}^{+0.097}$ & 8569$_{-707}^{+477}$ & 12.8$_{-0.6}^{+0.8}$ \\
J171529.75-732353.0 & \dots & He & 4810 $\pm$ 150 & 0.728$_{-0.085}^{+0.092}$ & 8.248$_{-0.092}^{+0.098}$ & 7275$_{-418}^{+257}$ & 12.8$_{-1.3}^{+1.7}$ \\
J171740.66+442805.1 & \dots & He & 4900 $\pm$ 70 & 0.483$_{-0.052}^{+0.055}$ & 7.853$_{-0.069}^{+0.068}$ & 4605$_{-708}^{+656}$ & \dots \\
J173059.01+115808.8 & \dots & H & 12950 $\pm$ 890 & 0.563$_{-0.043}^{+0.046}$ & 7.920$_{-0.057}^{+0.059}$ & 272$_{-53}^{+64}$ & 11.0$_{-1.9}^{+2.3}$ \\
J173059.01+115808.8 & \dots & He & 13610 $\pm$ 900 & 0.568$_{-0.058}^{+0.066}$ & 7.966$_{-0.074}^{+0.080}$ & 254$_{-52}^{+66}$ & 10.8$_{-2.5}^{+3.1}$ \\
J173149.47+033123.4 & \dots & H & 18590 $\pm$ 2250 & 0.253$_{-0.022}^{+0.027}$ & 6.944$_{-0.102}^{+0.113}$ & 762$_{946}^{+0}$ & \dots \\
J173149.47+033123.4 & \dots & He & 20130 $\pm$ 3670 & 0.261$_{-0.028}^{+0.030}$ & 7.071$_{-0.096}^{+0.099}$ & 33$_{-22}^{+28}$ & \dots \\
J174148.13+231434.1 & \dots & H & 5940 $\pm$ 90 & 0.526$_{-0.033}^{+0.035}$ & 7.895$_{-0.042}^{+0.042}$ & 1960$_{-121}^{+138}$ & 14.6$_{-1.7}^{+2.0}$ \\
J174148.13+231434.1 & \dots & He & 5810 $\pm$ 90 & 0.457$_{-0.034}^{+0.036}$ & 7.798$_{-0.047}^{+0.047}$ & 2052$_{-165}^{+199}$ & \dots \\
J174938.34+824717.4 & DA & H & 6840 $\pm$ 400 & 0.457$_{-0.059}^{+0.075}$ & 7.756$_{-0.087}^{+0.100}$ & 1159$_{-188}^{+243}$ & \dots \\
J180455.10+342800.8 & \dots & H & 9200 $\pm$ 320 & 0.360$_{-0.070}^{+0.084}$ & 7.519$_{-0.144}^{+0.137}$ & 448$_{-58}^{+69}$ & \dots \\
J180455.10+342800.8 & \dots & He & 9270 $\pm$ 420 & 0.321$_{-0.064}^{+0.085}$ & 7.465$_{-0.147}^{+0.146}$ & 436$_{-61}^{+73}$ & \dots \\
J182458.15+121300.1 & \dots & [H/He=$-$0.11] & 3370 $\pm$ 50 & 0.282$_{-0.020}^{+0.023}$ & 7.420$_{-0.041}^{+0.042}$ & 5239$_{-340}^{+368}$ & \dots \\
J185700.13-462723.1 & \dots & H & 12420 $\pm$ 1280 & 0.599$_{-0.069}^{+0.078}$ & 7.987$_{-0.089}^{+0.093}$ & 334$_{-89}^{+121}$ & 9.5$_{-2.4}^{+3.2}$ \\
J185700.13-462723.1 & \dots & He & 12140 $\pm$ 950 & 0.534$_{-0.077}^{+0.096}$ & 7.912$_{-0.106}^{+0.116}$ & 328$_{-76}^{+101}$ & 12.6$_{-4.0}^{+5.4}$ \\
J192615.75-462738.4 & \dots & H & 4200 $\pm$ 160 & 0.293$_{-0.064}^{+0.084}$ & 7.432$_{-0.135}^{+0.141}$ & 3396$_{-551}^{+942}$ & \dots \\
J192615.75-462738.4 & \dots & He & 4420 $\pm$ 80 & 0.351$_{-0.064}^{+0.073}$ & 7.588$_{-0.112}^{+0.110}$ & 3615$_{-597}^{+779}$ & \dots \\
J194000.62+834851.5 & \dots & H & 4780 $\pm$ 90 & 0.773$_{-0.037}^{+0.038}$ & 8.298$_{-0.039}^{+0.041}$ & 8834$_{-294}^{+273}$ & 13.5$_{-0.3}^{+0.4}$ \\
J194000.62+834851.5 & \dots & He & 4700 $\pm$ 80 & 0.715$_{-0.036}^{+0.040}$ & 8.229$_{-0.040}^{+0.042}$ & 7399$_{-179}^{+150}$ & 13.2$_{-0.6}^{+0.7}$ \\
J194111.82-180116.5 & \dots & H & 9020 $\pm$ 360 & 0.574$_{-0.080}^{+0.085}$ & 7.960$_{-0.104}^{+0.101}$ & 746$_{-114}^{+142}$ & 11.0$_{-2.9}^{+4.4}$ \\
J194111.82-180116.5 & \dots & He & 8840 $\pm$ 410 & 0.507$_{-0.080}^{+0.090}$ & 7.876$_{-0.111}^{+0.111}$ & 732$_{-123}^{+147}$ & 14.6$_{-4.5}^{+6.7}$ \\
J195039.59-585334.3 & \dots & H & 4600 $\pm$ 150 & 0.362$_{-0.074}^{+0.091}$ & 7.595$_{-0.129}^{+0.133}$ & 3295$_{-550}^{+1019}$ & \dots \\
J195039.59-585334.3 & \dots & He & 4670 $\pm$ 110 & 0.377$_{-0.072}^{+0.083}$ & 7.644$_{-0.119}^{+0.118}$ & 3350$_{-568}^{+905}$ & \dots \\
J200638.18+454451.6 & \dots & H & 4380 $\pm$ 80 & 0.482$_{-0.041}^{+0.044}$ & 7.836$_{-0.053}^{+0.055}$ & 6244$_{-673}^{+680}$ & \dots \\
J200638.18+454451.6 & \dots & He & 4460 $\pm$ 60 & 0.485$_{-0.037}^{+0.040}$ & 7.858$_{-0.049}^{+0.049}$ & 5601$_{-452}^{+443}$ & \dots \\
J201741.05+024051.9 & \dots & H & 11170 $\pm$ 430 & 0.409$_{-0.024}^{+0.026}$ & 7.618$_{-0.040}^{+0.042}$ & 287$_{-32}^{+36}$ & \dots \\
J201741.05+024051.9 & \dots & He & 11730 $\pm$ 570 & 0.382$_{-0.031}^{+0.036}$ & 7.599$_{-0.054}^{+0.059}$ & 253$_{-35}^{+41}$ & \dots \\
\hline
\end{tabular}
\end{table*}

\begin{table*}
\centering
\contcaption{}
\begin{tabular}{cccccccc}
\hline
Name & Spectral & Comp & $T_{\rm eff}$ & Mass & $\log{g}$ & Cooling Age & Total Age \\
     & Type     &      & (K)  & ($M_{\odot}$) & (cm s$^{-2}$) & (Myr) & (Gyr) \\
\hline
J202329.14+700151.2 & DA & H & 7200 $\pm$ 250 & 0.529$_{-0.045}^{+0.053}$ & 7.891$_{-0.060}^{+0.064}$ & 1213$_{-140}^{+162}$ & 13.7$_{-2.5}^{+2.8}$ \\
J203219.87+812459.5 & \dots & H & 9410 $\pm$ 410 & 0.393$_{-0.091}^{+0.110}$ & 7.597$_{-0.175}^{+0.165}$ & 449$_{-73}^{+97}$ & \dots \\
J203219.87+812459.5 & \dots & He & 9540 $\pm$ 530 & 0.356$_{-0.087}^{+0.112}$ & 7.552$_{-0.180}^{+0.174}$ & 429$_{-74}^{+99}$ & \dots \\
J203837.43-171815.4 & DA & H & 16970 $\pm$ 1380 & 0.537$_{-0.040}^{+0.048}$ & 7.854$_{-0.057}^{+0.063}$ & 99$_{-29}^{+39}$ & 12.2$_{-2.3}^{+2.4}$ \\
J204235.77-521820.1 & \dots & H & 5160 $\pm$ 110 & 0.561$_{-0.063}^{+0.066}$ & 7.967$_{-0.078}^{+0.077}$ & 4443$_{-1052}^{+1140}$ & 15.3$_{-1.5}^{+2.6}$ \\
J204235.77-521820.1 & \dots & He & 4990 $\pm$ 90 & 0.474$_{-0.058}^{+0.062}$ & 7.837$_{-0.080}^{+0.077}$ & 4247$_{-764}^{+749}$ & \dots \\
J205256.12+070929.5 & \dots & H & 3590 $\pm$ 150 & 0.429$_{-0.071}^{+0.081}$ & 7.741$_{-0.105}^{+0.105}$ & 7928$_{-1212}^{+1015}$ & \dots \\
J212742.15+154538.2 & \dots & H & 6090 $\pm$ 140 & 0.737$_{-0.039}^{+0.040}$ & 8.235$_{-0.043}^{+0.044}$ & 3758$_{-389}^{+389}$ & 9.1$_{-0.3}^{+0.4}$ \\
J212742.15+154538.2 & \dots & He & 5860 $\pm$ 130 & 0.635$_{-0.040}^{+0.042}$ & 8.102$_{-0.045}^{+0.047}$ & 3698$_{-527}^{+564}$ & 11.6$_{-0.7}^{+0.9}$ \\
J213951.50-255431.5 & \dots & H & 9080 $\pm$ 300 & 0.840$_{-0.056}^{+0.056}$ & 8.381$_{-0.063}^{+0.064}$ & 1466$_{-247}^{+307}$ & 5.2$_{-0.3}^{+0.5}$ \\
J213951.50-255431.5 & \dots & He & 8710 $\pm$ 330 & 0.754$_{-0.063}^{+0.064}$ & 8.277$_{-0.069}^{+0.070}$ & 1366$_{-218}^{+275}$ & 6.4$_{-0.7}^{+1.1}$ \\
J214959.88+540841.9 & DA & H & 10420 $\pm$ 270 & 0.484$_{-0.021}^{+0.023}$ & 7.786$_{-0.030}^{+0.032}$ & 418$_{-31}^{+33}$ & \dots \\
J220324.15+342044.5 & \dots & H & 4980 $\pm$ 120 & 0.352$_{-0.087}^{+0.103}$ & 7.565$_{-0.164}^{+0.152}$ & 2391$_{-475}^{+710}$ & \dots \\
J220324.15+342044.5 & \dots & He & 5010 $\pm$ 100 & 0.355$_{-0.089}^{+0.101}$ & 7.595$_{-0.163}^{+0.147}$ & 2580$_{-411}^{+791}$ & \dots \\
J222549.64+635748.2 & \dots & H & 4990 $\pm$ 170 & 0.468$_{-0.064}^{+0.081}$ & 7.806$_{-0.091}^{+0.101}$ & 3616$_{-935}^{+1211}$ & \dots \\
J222549.64+635748.2 & \dots & He & 4960 $\pm$ 140 & 0.449$_{-0.058}^{+0.069}$ & 7.789$_{-0.081}^{+0.089}$ & 3842$_{-723}^{+908}$ & \dots \\
J223707.28+063611.2 & \dots & H & 3860 $\pm$ 130 & 0.543$_{-0.089}^{+0.094}$ & 7.945$_{-0.112}^{+0.107}$ & 8911$_{-1040}^{+909}$ & 20.6$_{-3.0}^{+5.2}$ \\
J225755.79+294945.9 & \dots & H & 8340 $\pm$ 190 & 0.535$_{-0.033}^{+0.037}$ & 7.895$_{-0.044}^{+0.046}$ & 836$_{-64}^{+72}$ & 13.0$_{-1.8}^{+2.0}$ \\
J225755.79+294945.9 & \dots & He & 7960 $\pm$ 150 & 0.440$_{-0.030}^{+0.032}$ & 7.751$_{-0.044}^{+0.045}$ & 822$_{-55}^{+61}$ & \dots \\
J230853.31-134726.0 & \dots & H & 7420 $\pm$ 240 & 0.764$_{-0.057}^{+0.058}$ & 8.271$_{-0.064}^{+0.064}$ & 2176$_{-394}^{+448}$ & 7.0$_{-0.4}^{+0.7}$ \\
J230853.31-134726.0 & \dots & He & 7140 $\pm$ 230 & 0.676$_{-0.061}^{+0.063}$ & 8.162$_{-0.068}^{+0.069}$ & 1980$_{-333}^{+415}$ & 8.7$_{-1.0}^{+1.5}$ \\
J231908.89-061314.6 & DC & H & 4500 $\pm$ 70 & 0.446$_{-0.028}^{+0.031}$ & 7.770$_{-0.039}^{+0.040}$ & 5021$_{-494}^{+550}$ & \dots \\
J231908.89-061314.6 & DC & He & 4600 $\pm$ 50 & 0.461$_{-0.021}^{+0.022}$ & 7.813$_{-0.029}^{+0.029}$ & 4857$_{-291}^{+303}$ & \dots \\
J232410.45-592817.2 & DA & H & 10780 $\pm$ 310 & 0.531$_{-0.023}^{+0.025}$ & 7.874$_{-0.031}^{+0.032}$ & 424$_{-34}^{+37}$ & 12.8$_{-1.3}^{+1.4}$ \\
J232611.38-271448.4 & DC/DA: & H & 5320 $\pm$ 120 & 0.475$_{-0.063}^{+0.068}$ & 7.810$_{-0.088}^{+0.087}$ & 2440$_{-386}^{+636}$ & \dots \\
J235418.84-363405.6 & DA & H & 15140 $\pm$ 840 & 0.622$_{-0.032}^{+0.035}$ & 8.016$_{-0.040}^{+0.042}$ & 196$_{-33}^{+40}$ & 8.5$_{-1.1}^{+1.2}$ \\
J235435.58-322120.3 & DA & H & 10260 $\pm$ 710 & 0.534$_{-0.072}^{+0.094}$ & 7.882$_{-0.099}^{+0.116}$ & 487$_{-95}^{+124}$ & 12.7$_{-4.0}^{+4.9}$ \\
\hline
\end{tabular}
\end{table*}

We inspected the model fits for all 142 targets and decided on the atmospheric
composition based on the H$\alpha$ line profiles (if available), the quality of the fits for all of the photometric bands, and the
presence or absence of a near-infrared flux deficit due to molecular hydrogen. Table 2 presents the preferred composition, 
best-fitting effective temperature, mass, surface gravity, and the cooling age of each white dwarf. We cannot distinguish
between the pure H or pure He atmosphere solutions for 70 targets, as both sets of models provide acceptable fits to the
observed SEDs. We present both H and He atmosphere solutions for these 70 objects.

\subsection{Ultracool White Dwarfs}

The star formation history of the Galactic halo is best described by a single star burst model lasting over 1 Gyr with an
age of $\sim$12 Gyr \citep{reid05}. Hence, the majority of the intermediate-mass halo stars that have already evolved
into white dwarfs had enough time to cool down to temperatures below 4,000 K and become ultracool white dwarfs. 

Figure \ref{fig:ultra} shows the SEDs for six of the ultracool halo white dwarfs in our sample. Optical spectroscopy is
available for only two of these white dwarfs, and both are classified DC. However, all six should have featureless optical
spectra at these temperatures regardless of the atmospheric composition (unless they are metal-rich). All six of these
stars have optical and near-infrared
photometry (at least in the $J-$band) available, and the photometry clearly favors a H-rich composition for all but one of them.
The exception is J1503+5509, for which the photometry is not precise enough to distinguish between the H and He atmosphere
solutions. However, the remaining five stars in this figure show flux-deficits in the near-infrared compared to the He atmosphere
models, indicating the presence of H in their atmospheres. The best-fit temperatures range from 3590 K to 3940 for these five
white dwarfs.

The globular cluster M4 is currently forming $0.53 \pm 0.01 M_{\odot}$ white dwarfs \citep{kalirai12}. Hence, the Galactic halo is not old
enough to form lower mass white dwarfs ($M\leq0.5M_{\odot}$) from single stars. Therefore, low mass white dwarfs
are likely the result of enhanced mass loss in a binary system. Interestingly, only two of the ultracool white dwarfs shown
in Figure \ref{fig:ultra} have $M>0.5 M_{\odot}$, and therefore are consistent with single star evolution. J1049$-$7400 is a
H atmosphere white dwarf only 42 pc away from the Sun, with $T_{\rm eff} = 3940 \pm  180$  K,
$M = 0.694_{-0.069}^{+0.079} M_{\odot}$, and a cooling age of $10310_{-558}^{+480}$ Myr.
Similarly, J1503+5509 is a 0.51-0.52 $M_{\odot}$ white dwarf 75 pc away from the Sun, with $T_{\rm eff}=4000$-4190 K
and a cooling age of 6.7-7.9 Gyr, depending on its atmospheric composition. 

\section{Discussion}

\subsection{White Dwarf Cooling Ages}

Our detailed model atmosphere analysis indicates that 51 and 13 of our halo white dwarf candidates are
best explained by pure H and pure He atmosphere models, respectively. The latter include the DB white
dwarf J0007$-$3113. There are three DQ white dwarfs in our sample with He atmospheres and trace amounts
of C, and there are also five white dwarfs with mixed H/He atmospheres, including the ultracool white dwarfs
WD 0343+247 and J1824+1213. We cannot determine the atmospheric composition for the remaining 70 stars.

\begin{figure}
\vspace{-0.5in}
\includegraphics[width=\columnwidth]{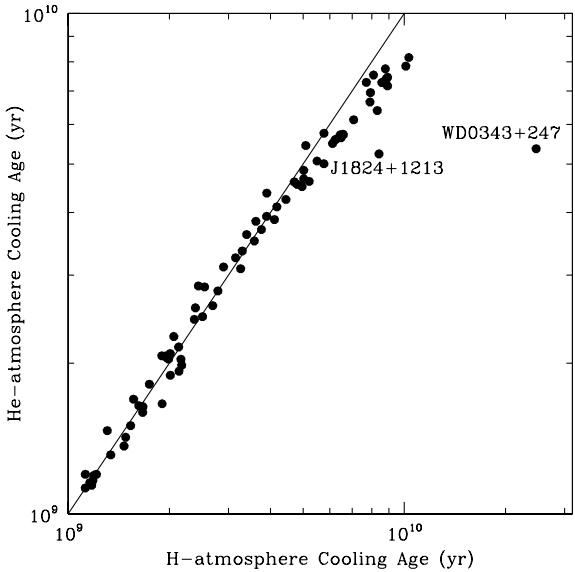}
\vspace{-0.8in}
\caption{Cooling ages of our halo white dwarf sample based on H- or He-atmosphere fits. Two objects with
mixed atmospheres, WD 0343+247 and J1824+1213, are labelled. The solid line marks the 1-1 ratio.}
\label{fig:comp}
\end{figure}

Figure \ref{fig:comp} shows the cooling ages of our targets based on H- and He-atmosphere model fits for ages
older than 1 Gyr. The solid line marks the 1-1 line. The white dwarf cooling ages of our targets range from 7 Myr
for J1657+2056 to 10.3 Gyr for J1049$-$7400. The choice of atmospheric composition has a small impact on
the cooling ages for stars hotter than about 5000 K \citep[e.g.,][]{harris06}. For example, J2203+3420 is either
a $T_{\rm eff}=4980 \pm  120$ K, pure H atmosphere white dwarf with a cooling age of 2391$_{-475}^{+710}$ Myr,
or a $T_{\rm eff}=5010 \pm  100$ K, pure He atmosphere white dwarf with a cooling age of 2580$_{-411}^{+791}$ Myr. 
However, the differences become significant for cooler white dwarfs, especially for cool white dwarfs that show
significant infrared flux deficits. For example, two of the mixed H/He atmosphere white dwarfs in our sample,
WD 0343+247 and J1824+1213, are significant outliers in Figure \ref{fig:comp}.

\begin{figure}
\vspace{-0.5in}
\includegraphics[width=\columnwidth]{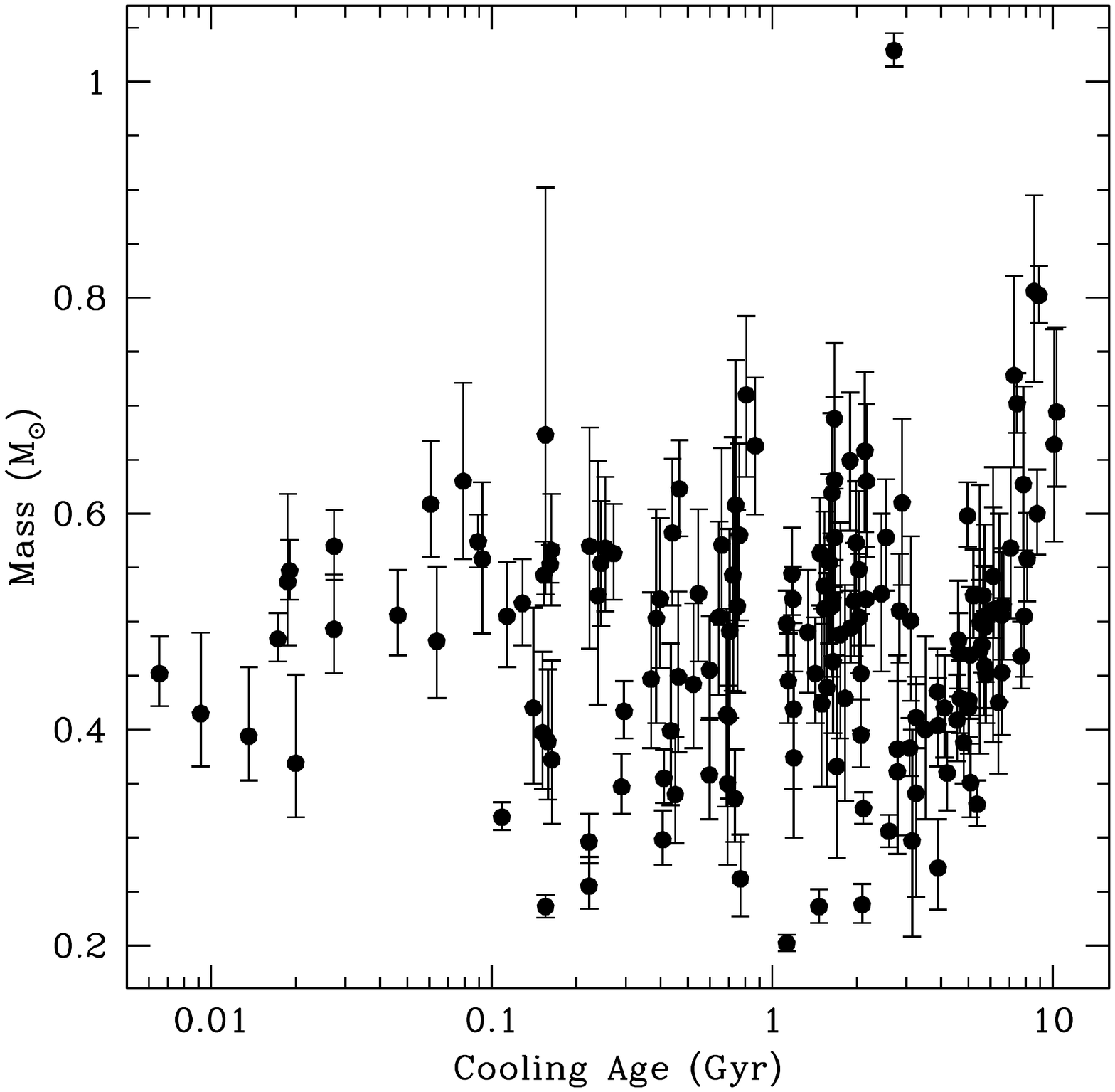}
\vspace{-0.8in}
\caption{Masses and cooling ages of our halo white dwarf sample. For objects with unknown atmospheric composition, both
H and He atmosphere solutions are shown. Many of the white dwarfs in our sample are low-mass ($M<0.5 M_{\odot}$), and
are likely formed in binary systems.}
\label{fig:massage}
\end{figure}

There are 11 white dwarfs with cooling ages that differ by more than 1 Gyr between the H and He atmosphere solutions; only 
five of these cannot be classified reliably based on the available data. The rest are clearly H-rich atmosphere
(either pure H or mixed H/He) white dwarfs. Hence, the unknown atmospheric composition does not have a significant
impact on our age measurements. Ignoring the pre-white dwarf evolutionary lifetimes, the coolest white dwarfs
in our sample present a firm lower limit of 10.3 Gyr for the age of the Galactic inner halo.

\subsection{Total Ages from Recently Formed White Dwarfs}

Figure \ref{fig:massage} shows the masses and cooling ages for our halo white dwarf sample. There are
a significant number of recently formed white dwarfs with relatively short
cooling timescales. Their total ages (cooling age + main-sequence + giant-branch evolutionary lifetimes) depend
largely on their main-sequence lifetimes, which are $\sim10$ Gyr for the progenitors of $0.5 M_{\odot}$ white dwarfs, but
significantly shorter for $0.6 M_{\odot}$ white dwarfs. Hence, a small uncertainty in the white dwarf mass can imply a large uncertainty
in the total ages.
Another complication in age measurements is that the initial-final mass relation is loosely constrained for $0.5 M_{\odot}$
white dwarfs, due to the difficulty of obtaining follow-up spectroscopy of similar white dwarfs in globular clusters.

\citet{kalirai12} used the recently formed white dwarfs in the 12.5 Gyr old globular cluster M4 and the results from
a {\em Hubble Space Telescope} imaging survey of 60 globular clusters \citep{sarajedini07} to derive a relation
that links the mass of remnants forming today to the parent population's age
\begin{equation}
\log ({\rm Age}) = \frac{\log(M_{\rm final} + 0.270) - 0.201}{-0.272} ~{\rm Gyr}, \label{kal}
\end{equation}
where $M_{\rm final}$ is the white dwarf mass in $M_{\odot}$. Since this relation is based on $>10$ Gyr old globular
clusters, it is valid over the white dwarf mass range of $\approx$0.5-0.6 $M_{\odot}$. Based on 4 halo white dwarfs,
\citet{kalirai12} used this relation to derive an age of $11.4 \pm 0.7$ Gyr for the Galactic inner halo, and
noted that this age measurement can be improved significantly by a larger sample of recently formed halo white dwarfs.

We use the same relation to estimate the pre-white dwarf evolutionary timescales and the total ages (including the
white dwarf cooling ages) of each source in our sample. These ages are included in Table 2. 
Ignoring the $M<0.5M_{\odot}$ white dwarfs, which are likely formed through binary evolution, all but one of the
white dwarfs with cooling ages $<1$ Gyr (Figure \ref{fig:massage}) have masses consistent with $M=0.5-0.6M_{\odot}$
within $1\sigma$. Similarly, only nine of the white dwarfs with $\geq1$ Gyr cooling ages have masses more than $1\sigma$ away
from this range. 

Two of the white dwarfs studied by \citet{kalirai12} are included in our halo white dwarf sample.
We measure a total age of $11.4^{+0.9}_{-0.8}$ Gyr for J1450+0733 (WD 1448+077), which is in excellent agreement
with \citet{kalirai12}. However, {\em Gaia} parallax measurements indicate a relatively low-mass of 
$M=0.493_{-0.041}^{+0.051} M_{\odot}$ for J1530$-$7505 \citep[WD 1524$-$749, see also][]{gianninas11}, and
we refrain from estimating its total age.

\begin{figure}
\vspace{-0.5in}
\includegraphics[width=\columnwidth]{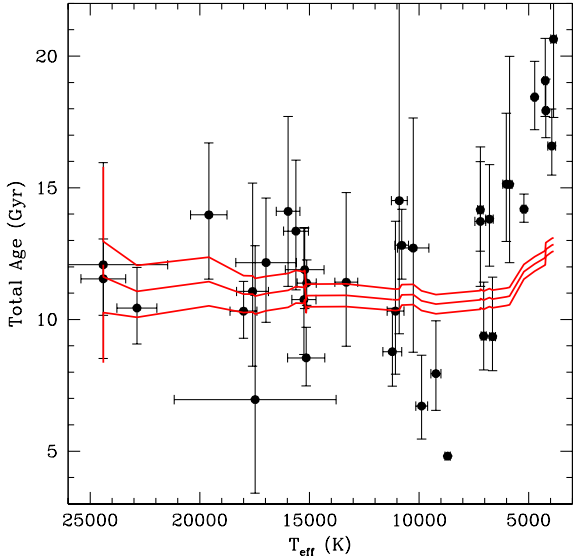}
\vspace{-0.8in}
\caption{Total ages, including the pre-white dwarf evolutionary lifetimes, of our halo white dwarf
sample. Only objects with $M\geq0.5 M_{\odot}$ and a preferred atmospheric composition are shown. The red lines show
the cumulative average and its 2$\sigma$ error range as a function of decreasing effective temperature.}
\label{fig:total}
\end{figure}

Figure \ref{fig:total} shows the total ages of our halo white dwarf sample as a function of effective temperature. 
Only objects more massive than $0.5M_{\odot}$ and with a preferred atmospheric composition are shown. 
The red lines show the cumulative average and its 2$\sigma$ error range as a function of decreasing effective temperature.
Limiting our sample to the recently formed white dwarfs with $T_{\rm eff}>10,000$ K, or with white dwarf cooling ages of
$<500$ Myr, the weighted mean age is $10.9 \pm 0.4$ Gyr, which is again in excellent agreement with
\citet{kalirai12}'s estimate of $11.4 \pm 0.7$ Gyr. Including the objects with unknown composition, but
with $T_{\rm eff}>10,000$ K, and using either the H or He solution still gives the same mean age of
$10.9 \pm 0.4$ Gyr. However, this method clearly over-estimates the total ages for the coolest white dwarfs in
our sample, as they all have estimated total ages above 15 Gyr.

\subsection{Total Ages from the Coolest White Dwarfs}

The age versus final mass relation derived by \citet{kalirai12} is relatively flat in the $0.5-0.6 M_{\odot}$ range, giving
main-sequence lifetimes of 9-14 Gyr. However, this relation is based on spectroscopy of relatively bright white dwarfs
in globular clusters; it is not calibrated for the faintest white dwarfs with relatively
long white dwarf cooling ages. \citet{hansen07} find that the truncation in the white dwarf luminosity function of the globular
cluster NGC 6397 occurs when the mass of the main-sequence progenitors significantly increase, which corresponds
to a significant decrease in age. For example, J1049$-$7400 has the largest white dwarf cooling age in our sample, but
it is relatively massive with $M=0.694_{-0.069}^{+0.079} M_{\odot}$. Hence, the age-mass relation given in equation \ref{kal}
is not appropriate in this case.

We use the initial final mass relation from \citet{kalirai09}, and the pre-white dwarf evolutionary lifetime of
halo-metallicity main-sequence stars with [Fe/H]$=-1.5$ from \citet{hurley00} to estimate the total ages of the
coolest white dwarfs in our sample. Table 3 presents the total ages based on this method for five of our halo
white dwarfs with $\geq5$ Gyr cooling ages and with a preferred atmospheric composition. 

\begin{table}
\centering
\caption{Total ages for five of our halo white dwarfs with $M>0.5 M_{\odot}$, a preferred atmospheric composition, and
relatively long white dwarf cooling age of $\geq5$ Gyr.  Here we use the initial final mass relation from \citet{kalirai09},
and the pre-white dwarf evolutionary lifetime of halo-metallicity main-sequence stars from \citet{hurley00} to estimate the total age.}
\begin{tabular}{ccc}
\hline
Name & Composition &  Total Age \\
          &                      &       (Gyr) \\
\hline
J075015.56$+$071109.3 & He & $\geq 10.2$ \\
J104957.54$-$740028.4  & H   & $10.9^{+0.2}_{-0.0}$ \\
J114730.20$-$745738.2  & H   & $10.4^{+1.3}_{-0.2}$ \\
J131253.16$-$472808.9  & H   &  $11.8^{+\dots}_{-1.6}$ \\
J223707.28$+$063611.2 & H    & $14.4^{+\dots}_{-3.5}$ \\
\hline
\end{tabular}
\end{table}

The total ages for these five white dwarfs range from $\geq10.2$ Gyr to 14.4 Gyr. For some of these targets we are only
able to put a lower limit on the total age since the mass is near the boundary for low-mass white dwarfs ($0.5 M_{\odot}$).
Regardless of this, there are two
white dwarfs with well constrained ages. J1049$-$7400 is a $3940 \pm 180$ K, H atmosphere white dwarf with a best-fit
total age of 10.9 Gyr. Taking into account the errors in mass and cooling age, the total age is constrained to between 10.9
and 11.1 Gyr,  providing an excellent age measurement of the inner halo. Similarly, J1147$-$7457 is a $4210 \pm 80$ K,
H atmosphere white dwarf with a best-fit total age of 10.4 Gyr, and an age range of 10.2-11.7 Gyr. These ages are consistent
with the age constraints from the recently formed white dwarfs discussed above. 

Our conservative selection excluded some of the well known cool white dwarfs like J1102+4113 \citep{hall08}.
{\em Gaia} Data Release 2 provided a parallax measurement of $\varpi = 28.649 \pm 0.289$ mas for this white dwarf, which
is consistent with the distance measurement reported in \citet[][$33.7 \pm 2.0$ pc]{kilic12}. J1102+4113 has a total age
of 10.6-11.1 Gyr, which is similar to the total ages for the coolest white dwarfs in our sample.

\section{Conclusions}

We use the {\em Gaia} $\varpi \geq 5 \sigma_{\varpi}$ white dwarf sample to measure the velocity dispersion of the Galactic disc,
and identify 142 halo white dwarfs in the solar neighborhood. We perform a detailed model
atmosphere analysis of these white dwarfs and constrain their masses and cooling ages. We estimate an inner halo age of
$10.9 \pm 0.4$ Gyr from the recently formed white dwarfs found in our sample, and the coolest white dwarfs
also confirm this age measurement.

\begin{figure}
\vspace{-0.5in}
\includegraphics[width=\columnwidth]{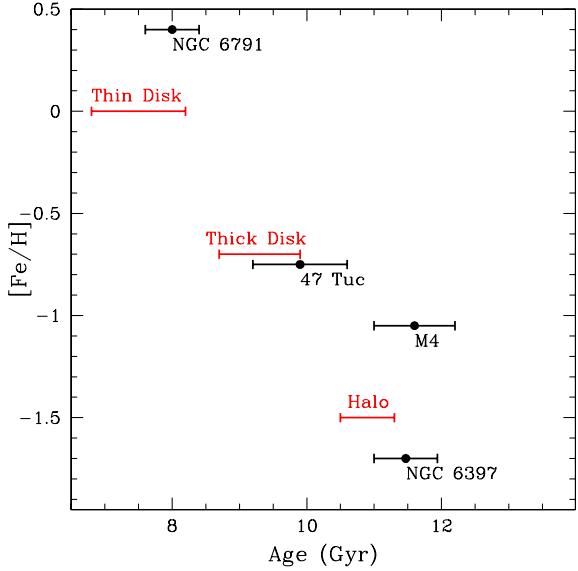}
\vspace{-0.8in}
\caption{Age-Metallicity relation based on white dwarfs in the open cluster NGC 6791, globular clusters 47 Tuc,
M4, and NGC 6397 \citep[][and references therein]{hansen13}, the local thin disk and thick disk \citep{kilic17}, and
halo (this study).}
\label{fig:ages}
\end{figure}

Figure \ref{fig:ages} compares our halo age measurement with the white dwarf luminosity function based ages of the
local thin disc and thick disc populations \citep{kilic17}, as well as four of the oldest clusters observed
by the {\em Hubble Space Telescope}. \citet{hansen07,hansen13} measured ages of 11.5 $\pm$ 0.5
and 9.9 $\pm$ 0.7 Gyr for the globular clusters NGC 6397 and 47 Tuc, respectively. Similarly, \citet{bedin09}
and \citet{garciaberro10} measured ages of $11.6 \pm 0.6$ Gyr and $8.0 \pm 0.4$ Gyr for the globular cluster M4
and the metal-rich open cluster NGC 6791, respectively. 

Our halo age estimate is consistent with the white dwarf based ages of the globular clusters M4, NGC 6397, and 47 Tuc,
as well as the age measurements for the inner halo from \citet[][$11.4\pm0.7$ Gyr, based on 4 white dwarfs]{kalirai12},
and \citet[][$12.11^{+0.85}_{-0.86}$ Gyr, based on 4 white dwarfs with parallax measurements]{si17}. 
Note that all of the errors quoted here are internal errors. Studying a
large number of spectroscopically confirmed blue horizontal branch stars, \citet{carollo16} find a clear concentration of
very old stars extending out to 10-15 kpc from the Galactic center and measure an age gradient of $-25 \pm 1$ Myr
kpc$^{-1}$ as a function of radial Galactocentric distance, $R$. They measure an age of $\approx$11.4 Gyr for $R=8$ kpc,
in the Sun's vicinity. This is consistent with our white dwarf based age measurements, confirming that the theoretical
uncertainties due to the unknown core composition, helium layer mass, crystallization and phase separation are
$\sim0.5$ Gyr \citep{montgomery99} for the oldest white dwarfs in the Galaxy.

\section*{Acknowledgements}

We thank the referee, H. Richer, for a constructive report.
This work is supported in part by the NSERC Canada and by the Fund FRQ-NT (Qu\'ebec).

This work presents results from the European Space Agency (ESA) space mission Gaia.
Gaia data are being processed by the Gaia Data Processing and Analysis Consortium (DPAC).
Funding for the DPAC is provided by national institutions, in particular the institutions
participating in the Gaia MultiLateral Agreement (MLA).
The Gaia mission website is https://www.cosmos.esa.int/gaia.
The Gaia archive website is https://archives.esac.esa.int/gaia.

This work is based in part on data obtained as part of the UKIRT Infrared Deep Sky Survey and
the UKIRT Hemisphere Survey (UHS). The UHS is a partnership between the UK STFC, The
University of Hawaii, The University of Arizona, Lockheed Martin, and NASA.

The Pan-STARRS1 Surveys (PS1) have been made possible through contributions of the Institute for Astronomy, the University of Hawaii, the Pan-STARRS Project Office, the Max-Planck Society and its participating institutes, the Max Planck Institute for Astronomy, Heidelberg and the Max Planck Institute for Extraterrestrial Physics, Garching, The Johns Hopkins University, Durham University, the University of Edinburgh, Queen's University Belfast, the Harvard-Smithsonian Center for Astrophysics, the Las Cumbres Observatory Global Telescope Network Incorporated, the National Central University of Taiwan, the Space Telescope Science Institute, the National Aeronautics and Space Administration under Grant No. NNX08AR22G issued through the Planetary Science Division of the NASA Science Mission Directorate, the National Science Foundation under Grant No. AST-1238877, the University of Maryland, and Eotvos Lorand University (ELTE).

Funding for SDSS-III has been provided by the Alfred P. Sloan Foundation, the Participating Institutions, the National Science Foundation, and the U.S. Department of Energy Office of Science. The SDSS-III web site is http://www.sdss3.org/.
SDSS-III is managed by the Astrophysical Research Consortium for the Participating Institutions of the SDSS-III Collaboration including the University of Arizona, the Brazilian Participation Group, Brookhaven National Laboratory, Carnegie Mellon University, University of Florida, the French Participation Group, the German Participation Group, Harvard University, the Instituto de Astrofisica de Canarias, the Michigan State/Notre Dame/JINA Participation Group, Johns Hopkins University, Lawrence Berkeley National Laboratory, Max Planck Institute for Astrophysics, Max Planck Institute for Extraterrestrial Physics, New Mexico State University, New York University, Ohio State University, Pennsylvania State University, University of Portsmouth, Princeton University, the Spanish Participation Group, University of Tokyo, University of Utah, Vanderbilt University, University of Virginia, University of Washington, and Yale University.

\bibliographystyle{mnras}
\bibliography{master}

\bsp
\label{lastpage}

\end{document}